## Title

- A physics-informed foundation model for quantitative diffusion MRI

## Authors

Zihan Li[1]†, Jialan Zheng[1]†, Ziyu Li[2*], Xun Yuan[3], Kasidit Anmahapong[1], Ziang Wang[1], Mingxuan Liu[1], Hongjia Yang[1], Yifei Chen[1], Zhuhao Wang[1], Yuhang He[1], Fang Chen[4], Rui Li[1], Huaiqiang Sun[5], Yi Liao[3], Congyu Liao[6], Yang Yang[6], Haibo Qu[3], Xue Zhang[7], Hongen Liao[1*], Qiyuan Tian[1*]

for the Alzheimer's Disease Neuroimaging Initiative[#]

## Affiliations

[1]School of Biomedical Engineering, Tsinghua University, Beijing, China.

[2]Oxford Centre for Integrative Neuroimaging, FMRIB, Nuffield Department of Clinical Neurosciences, University of Oxford, Oxford, United Kingdom.

[3]Department of Radiology, West China Second University Hospital, Sichuan University, Chengdu, China.

[4]School of Biomedical Engineering and the Institute of Medical Robotics, Shanghai Jiaotong University, Shanghai, China.

[5]Department of Radiology, Institution of Radiology and Medical Imaging, West China Hospital, Sichuan University, Chengdu, China.

[6]Department of Radiology and Biomedical Imaging, University of California San Francisco, San Francisco, USA.

[7]Department of Psychiatry and Behavioral Sciences, Stanford University School of Medicine, Stanford, USA.

†These authors contributed equally to this work
*Corresponding to:
Qiyuan Tian, Ph.D., qiyuantian@tsinghua.edu.cn
Hongen Liao, Ph.D., liao@tsinghua.edu.cn
Ziyu Li, Ph.D., ziyu.li@ndcn.ox.ac.uk

[#]Data used in preparation of this article were obtained from the Alzheimer's Disease Neuroimaging Initiative (ADNI) database (adni.loni.usc.edu). As such, the investigators within the ADNI contributed to the design and implementation of ADNI and/or provided data but did not participate in analysis or writing of this report. A complete listing of ADNI investigators can be found at: http://adni.loni.usc.edu/wp-content/uploads/how_to_apply/ADNI_Acknowledgement_List.pdf.

**Abstract**

Understanding the human brain requires access to its microscopic tissue architecture, the very substrate of neural communication, development, and disease. Diffusion magnetic resonance imaging (MRI) provides the only noninvasive window into whole-brain microstructure *in vivo*, yet reliable quantitative microstructural mapping remains confined to specialized research settings, requiring dense diffusion sampling and carefully optimized acquisition protocols. To address this gap, we present a physics-informed generative microstructure network (PIGMENT) that learns a universal generative prior of human brain microstructure and adapts it in a zero-shot manner to each participant's measured diffusion data to recover subject-specific maps across diverse acquisition protocols. Trained on up to 11,375 scans spanning multiple sites, vendors, field strengths, and acquisition protocols, PIGMENT enabled reliable quantitative microstructural mapping for tensor, kurtosis, and NODDI models across external datasets from five independent centers, despite substantial variation in acquisition protocols. It remained effective in regimes where conventional fitting becomes unreliable, recovering quantitatively meaningful microstructural maps from extremely sparse acquisitions (as few as three diffusion-weighted images), while supporting downstream tractography and structural connectivity mapping. The estimates demonstrated strong biological validity by preserving informative signals, including submillimeter cortical microarchitectural patterns and early-childhood white matter developmental trajectories from 10-fold accelerated scans. Furthermore, under constrained scanning conditions, PIGMENT enables reliable quantitative tensor mapping on cost-efficient low-field systems and the extraction of tumor-related microstructural biomarkers and peritumoral fiber pathways using ultra-fast clinical protocols. Together, these results establish PIGMENT as a physics-informed foundation model that extends quantitative diffusion MRI into acquisition regimes that are traditionally too sparse, heterogeneous, or clinically constrained for reliable microstructural analysis.

## MAIN TEXT

## INTRODUCTION

The human brain's functional capacity, developmental trajectory and vulnerability to disease are all rooted in its microscopic tissue microarchitecture, making its *in vivo* measurement fundamentally important[1,2]. A typical human brain contains approximately 86 billion neurons[3] interconnected by nearly 160,000 kilometers of myelinated axons[4]. Their organization shapes how signals propagate within and across brain regions, supports development and plasticity across the lifespan, and is disrupted in a wide range of neurological and psychiatric conditions[5-8]. Characterizing brain microstructure *in vivo* therefore matters not only for understanding fundamental principles of brain organization, but also for identifying biologically grounded markers that can dynamically track disease and treatment response[9-11].

Diffusion magnetic resonance imaging (MRI) stands as the only noninvasive *in vivo* window into whole-brain tissue microstructure and structural connectivity, making it an indispensable tool for systems neuroscience. By modeling the microscopic diffusion of water molecules, diffusion MRI translates invisible cellular barriers into quantitative maps of tissue organization[12]. These readouts capture the local complexity of neural tissue and can be seamlessly integrated with tractography to reconstruct the brain's global wiring diagram[13,14]. By linking microscopic cellular properties to large-scale structural connectivity, quantitative diffusion MRI elegantly translates between local tissue architecture and systems-level brain phenotypes[13,15].

Despite its promise, robust quantitative microstructural imaging relies on stringent imaging regimes to ensure both dense sampling and high signal quality. Conventional voxel-wise fitting amplifies noise or yields physically implausible parameter maps when data are insufficient. Consequently, a fundamental conflict emerges between acquisition feasibility and estimation fidelity. In practice, these stringent requirements form a major bottleneck in scenarios constrained by limited scan time, low field strength, poor patient compliance, or rigid clinical routines. The challenge is particularly acute in pediatric imaging[16,17], low-field imaging[18,19], submillimeter acquisitions[20,21], and time-constrained clinical protocols[22], where conventional model fitting becomes unstable or entirely intractable. As a result, much of the potential value of diffusion MRI remains inaccessible outside specialized research environments.

Deep learning techniques have emerged as an effective strategy for accelerating diffusion MRI data acquisition and improving quantitative mapping from sparse data. Prior work has shown that neural networks can recover diffusion tensor metrics and even advanced microstructural estimates from under-sampled acquisitions. Network architectures in this domain have evolved significantly, progressing from early multi-layer perceptrons (e.g., q-DL) that process signals in a voxel-wise manner[23-25], to convolutional neural networks (e.g., DeepDTI) designed to exploit spatial neighborhood correlations for noise suppression[26,27], and most recently, to transformers capable of capturing long-range spatial dependencies[28]. For instance, these studies have demonstrated the feasibility of recovering accurate diffusion tensor model metrics from as few as six diffusion-weighted images (DWIs)[26], and high-fidelity neurite orientation dispersion and density imaging (NODDI) maps from just 12 directions per b-value[28]. However, most approaches remain tied to protocol-specific training distributions or supervised mappings to reference measurements. As a result, their performance often degrades or fails entirely when acquisition design, scanner characteristics, spatial resolution, or diffusion encoding differ from those seen during training.

These limitations highlight the need for approaches that can flexibly accommodate arbitrary diffusion inputs while remaining robust across heterogeneous acquisition protocols and imaging environments. One strategy has been to derive more unified input representations, such as spherical harmonic (SH) coefficients (e.g., SH-DTI[29]), to improve generalization across gradient directions[30,31]. However, this approach remains sensitive to scanner- and protocol-dependent variation and still requires a minimum sampling density for reliable computation. A complementary direction is self-supervised, physics-informed modeling. In our previous work, DIMOND established a flexible framework for microstructural estimation directly from arbitrary diffusion encodings by integrating biophysical forward models into the optimization process[32]. More broadly, recent progress in medical foundation models suggests an opportunity to extend this direction further by learning large-scale, transferable priors from heterogeneous imaging data[33-35]. For diffusion MRI, such priors could provide a principled way to improve robustness across diverse acquisition designs while retaining compatibility with subject-specific measured signals.

In this study, we present PIGMENT (physics-informed generative microstructure network), a foundation model for zero-shot quantitative diffusion MRI from arbitrary acquisition protocols. PIGMENT learns a universal generative prior of human brain microstructure from large-scale, heterogeneous multi-center data and then adapts this prior to each participant's measured diffusion data through physics-informed inference to recover subject-specific microstructural maps. We evaluated PIGMENT across three complementary dimensions: methodological generalizability, by testing reconstruction fidelity across biophysical models, diffusion encodings, and acquisition protocols; biological validity, by assessing submillimeter cortical microarchitectural features and early-childhood white matter developmental trajectories; and translational utility, by examining quantitative mapping under low-field and highly accelerated neuro-oncology protocols. We show that this foundation model enables reliable quantitative diffusion MRI in regimes that are too sparse, heterogeneous, or clinically constrained for conventional microstructural analysis.

## RESULTS

### Mapping human brain microstructure using a generative prior

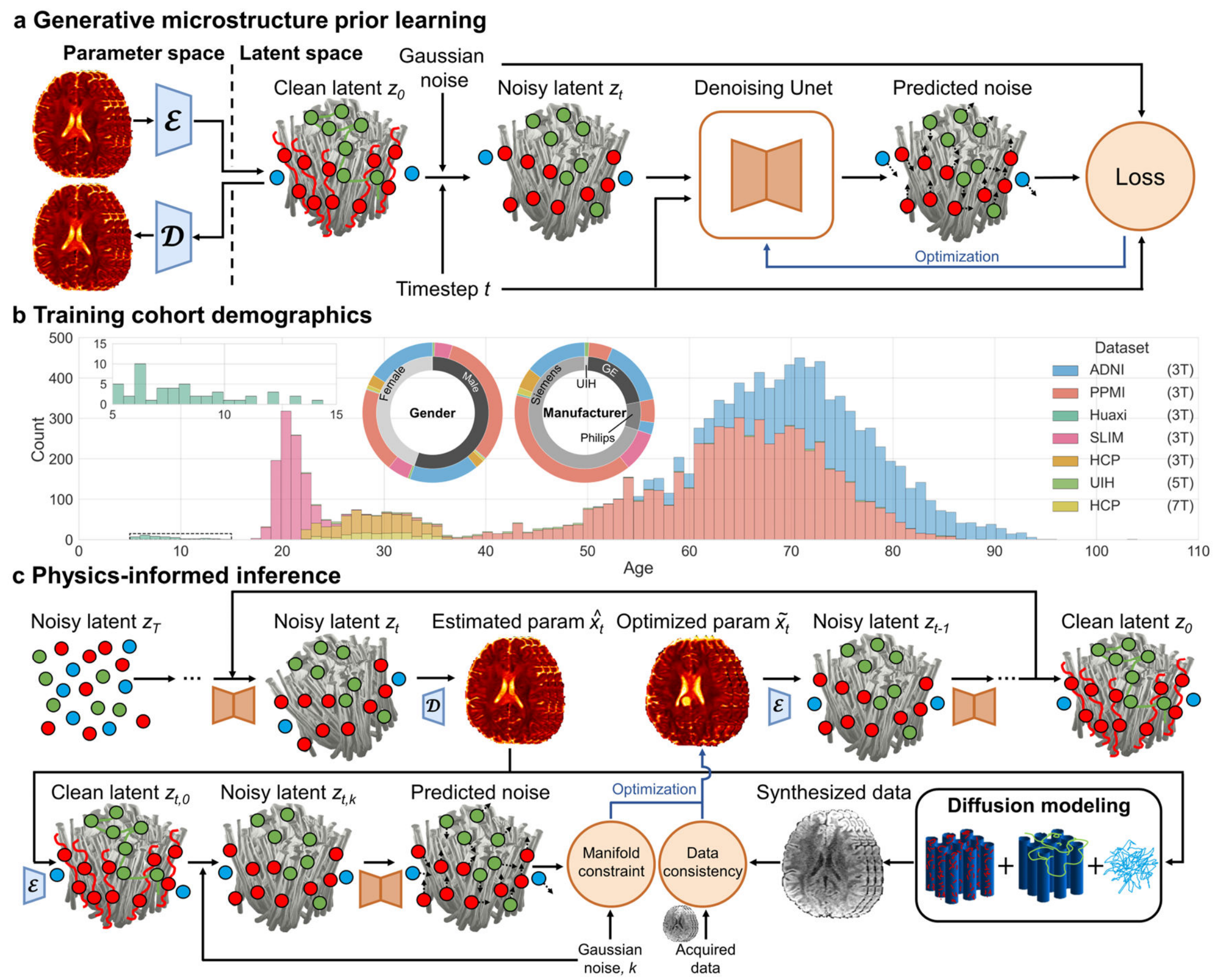

**Figure 1 | PIGMENT framework. a**, Generative microstructure prior learning. PIGMENT learns a comprehensive prior of human brain microstructure from large-scale data, embedding it into a microstructure latent diffusion model that integrates an autoencoder with a denoising U-Net. The autoencoder establishes a bidirectional mapping between the parameter space and the latent space through an encoder ($\mathcal{E}$) and a decoder ($\mathcal{D}$). The denoising U-Net predicts the noise component added to the latent representation at each diffusion time step $t = 1, \ldots, T$. **b**, Training cohort demographics. The large-scale cohort includes 11,375 scans from subjects aged 5–104 years, with a balanced sex ratio (45% female, 55% male) and broad representation across scanner vendors (Siemens 70%, General Electric 21%, Philips 8%, United Imaging Healthcare 1%) and field strengths (3, 5, and 7 Tesla). **c**, Physics-informed inference. PIGMENT generates individual-specific microstructure by alternately denoising the latent microstructure representation with the pre-trained latent diffusion model and refining it via dual-constraint optimization that enforces data and manifold consistency.

We developed PIGMENT as a physics-informed foundation model for quantitative diffusion MRI that combines a learned generative prior of human brain microstructure (Fig. 1a), trained across a large-scale and heterogeneous dataset (Fig. 1b), with subject-specific physics-informed inference (Fig. 1c). Specifically, the prior learning module establishes a comprehensive, high-quality prior of human brain microstructure by embedding a large-scale dataset of 11,375 scans into a latent diffusion model. This diverse training cohort features a relatively balanced sex distribution (45% female, 55% male), broad scanner-vendor coverage (Siemens 70%, General Electric 21%, Philips 8%, United Imaging Healthcare 1%), and varying field strengths (3 Tesla, 5 Tesla, and 7 Tesla) (Fig. 1b).

We demonstrated that the learned prior embeds rich intrinsic brain microstructural information. For few-shot multimodal image synthesis, a decoder trained on features extracted from a single b = 0 image and six DWIs from only five Chinese Human Connectome Project (CHCP)[36] subjects generated high-quality T1-weighted (T1w, Fig. S1), T2-weighted (T2w, Fig. S2), and diffusion-weighted images (DWI, Fig. S3), and enabled accurate brain segmentation from the synthesized T1w images (Fig. S4), substantially outperforming the benchmark image translation network Pix2Pix[37] trained on those raw images (Fig. S5). Collectively, these results demonstrate that the learned diffusion MRI prior provides a strong representational basis, which underpins PIGMENT's robust generalization toward high-fidelity quantitative brain imaging.

Subsequently, to ensure accurate subject-specific estimation across heterogeneous acquisition protocols, the physics-informed inference module refines the learned prior using subject-specific acquired data to generate individualized microstructural maps. Specifically, a dual-constraint optimization strategy is incorporated into the generative sampling trajectory. This strategy iteratively enforces consistency between model-predicted signals derived from the estimated microstructure parameters and the acquired measurements, while simultaneously constraining the solution within the valid manifold of the learned prior. In the absence of dual-constraint optimization, PIGMENT generated slightly blurred but anatomically plausible individualized microstructural maps from a single b = 0 image (Fig. S6), highlighting both the representational strength of the learned generative prior and the necessity of subject-specific physics-informed optimization.

The efficacy of PIGMENT was systematically evaluated using out-of-distribution (OOD) data from two public datasets and clinical or research cohorts across five independent centers in China (*n*=3 centers), the United States (*n*=1 center), and the United Kingdom (*n*=1 center).

## Enabling robust generation across biophysical models and protocols

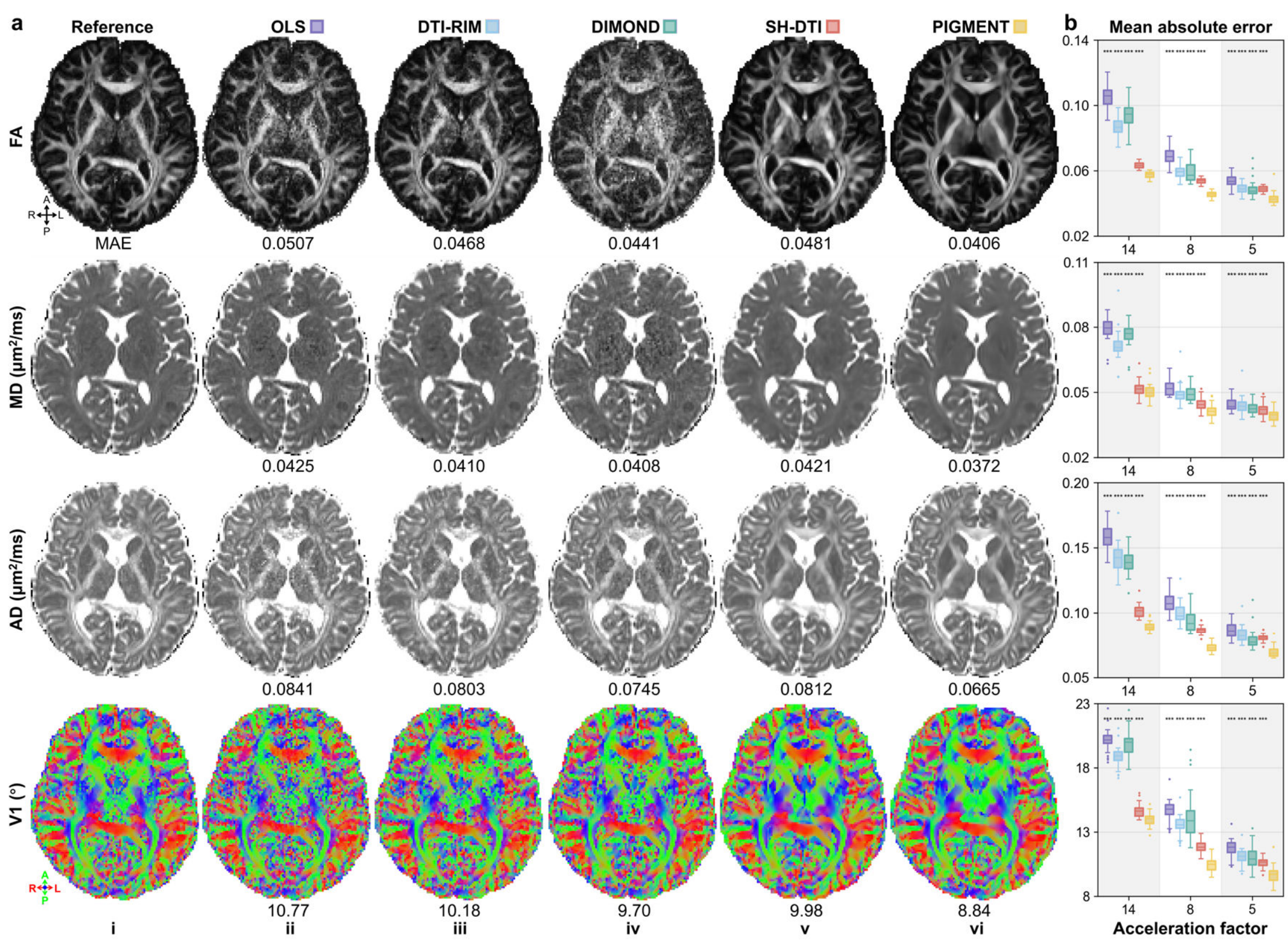

**Figure 2 | Robustness across diffusion encodings for diffusion tensor imaging. a**, Exemplar axial maps of diffusion tensor model metrics, including fractional anisotropy (FA), mean diffusivity (MD), axial diffusivity (AD), and primary eigenvector (V1), are shown for a representative subject. Maps were generated using OLS from the full single-shell data of this CHCP subject (reference, i) and using other methods from accelerated data subsets with 5-fold acceleration (ii-vi). **b**, Mean absolute errors (MAEs) of FA, MD, AD, and V1 relative to the reference within white matter are shown for 21 CHCP subjects across acceleration factors of 14, 8, and 5. Box plots illustrate the distribution of metric values, where the midline represents the median, box edges indicate the first and third quartiles, and whiskers extend 1.5 times the interquartile range (IQR) with outliers plotted individually. Asterisks indicate the significance level of differences between each method and PIGMENT (two-sided paired *t*-test; $^{***}P < 0.001$, $^{**}P < 0.01$, $^{*}P < 0.05$).

PIGMENT achieves robust reconstruction accuracy across microstructural models, diffusion encodings, and acceleration levels. We evaluated the performance of PIGMENT across three widely adopted microstructural models (tensor model, kurtosis model, and NODDI model) using the CHCP dataset. To ensure broad applicability, our evaluations encompassed diverse diffusion encoding schemes across varying acceleration factors. Furthermore, the testing cohort was highly representative, spanning a wide adult age range (19–79 years) with a balanced sex distribution (10 males, 11 females).

PIGMENT consistently yielded spatially coherent, high-quality tensor metrics that closely approximated the reference across all acceleration factors (Fig. 2a). Specifically, SH-DTI excelled at high acceleration (14-fold) owing to robust prior regularization, whereas DIMOND performed better at lower acceleration (5-fold) by leveraging subject-specific signal characteristics. PIGMENT seamlessly integrated these conceptual advantages, achieving superior fidelity and the lowest error rates across the entire acceleration spectrum.

Quantitatively, PIGMENT demonstrated substantial performance improvements relative to the standard ordinary least-squares (OLS) regression and significantly outperformed all other competing methods ($P < 0.05$, two-sided paired $t$-test). Specifically, for fractional anisotropy (FA), PIGMENT reduced the mean absolute error (MAE) by 45%, 34%, and 10% at 14-fold, 8-fold, and 5-fold acceleration, respectively (Fig. 2b, Fig. S7). In contrast, methods relying solely on priors, such as SH-DTI, yielded higher errors than OLS, particularly for mean diffusivity (MD) at lower acceleration factors. Furthermore, structural connectivity matrices derived from PIGMENT using only six DWIs demonstrated high consistency with the reference (Fig. S8; concordance correlation coefficient (CCC) = 0.983, $R = 0.985$).

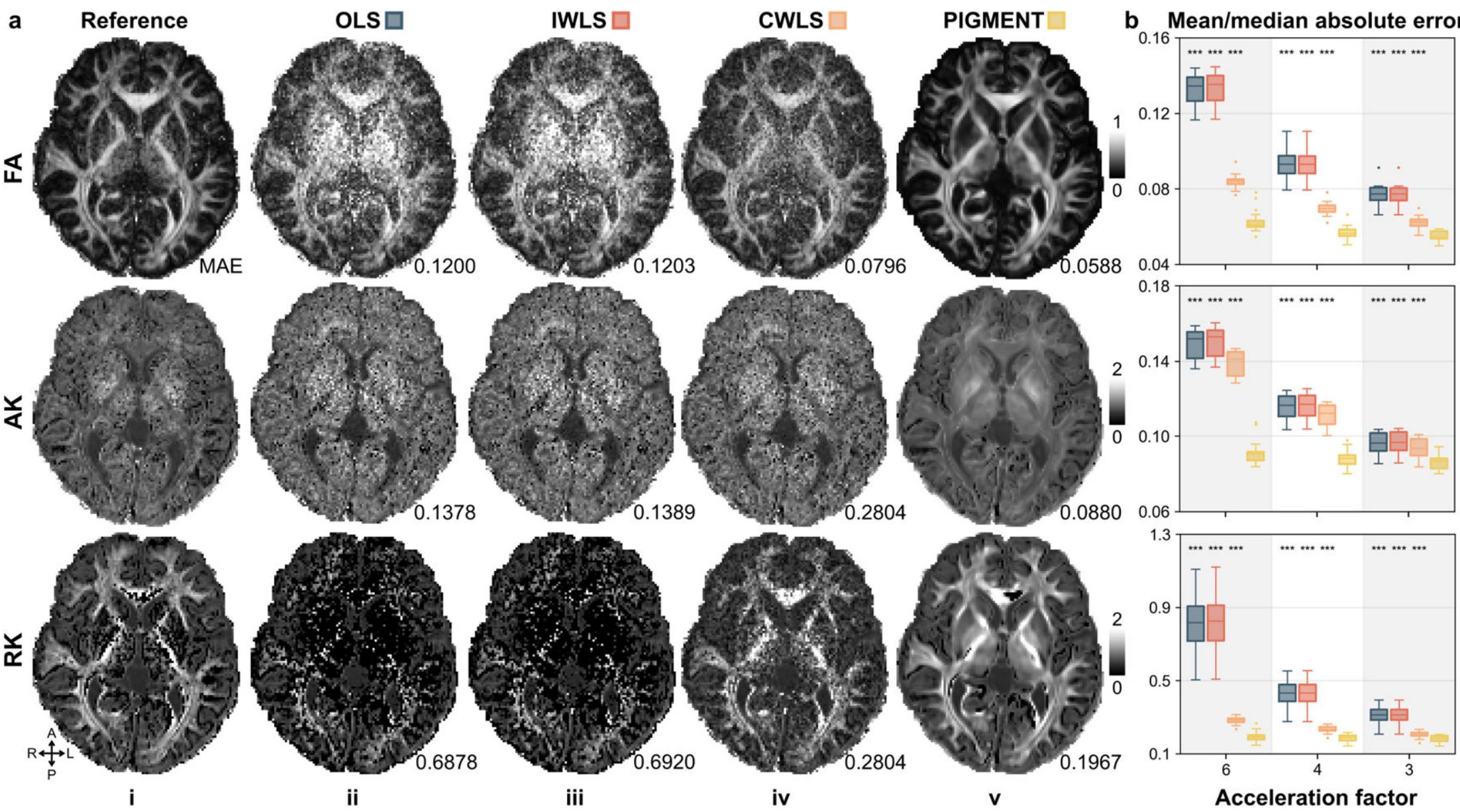

**Figure 3 | Robustness across diffusion encodings for diffusion kurtosis imaging. a**, Exemplar axial maps of diffusion kurtosis model metrics, including fractional anisotropy (FA), axial kurtosis (AK), and radial kurtosis (RK), are shown for a representative subject. Maps were generated using OLS over the full acquisition data (reference, i) and using other methods from accelerated data subsets with 6-fold accelerations (ii-v). **b**, Mean absolute errors of FA, and median absolute errors (MAEs) of AK, and RK relative to the reference within white matter are shown for 21 CHCP subjects across acceleration factors of 6, 4, and 3. Box plots illustrate the distribution of metric values, where the midline represents the median, box edges indicate the first and third quartiles, and whiskers extend 1.5 times the interquartile range (IQR) with outliers plotted individually. Asterisks indicate the significance level of differences between each method and PIGMENT (two-sided paired $t$-test; $^{***}P < 0.001$, $^{**}P < 0.01$, $^{*}P < 0.05$).

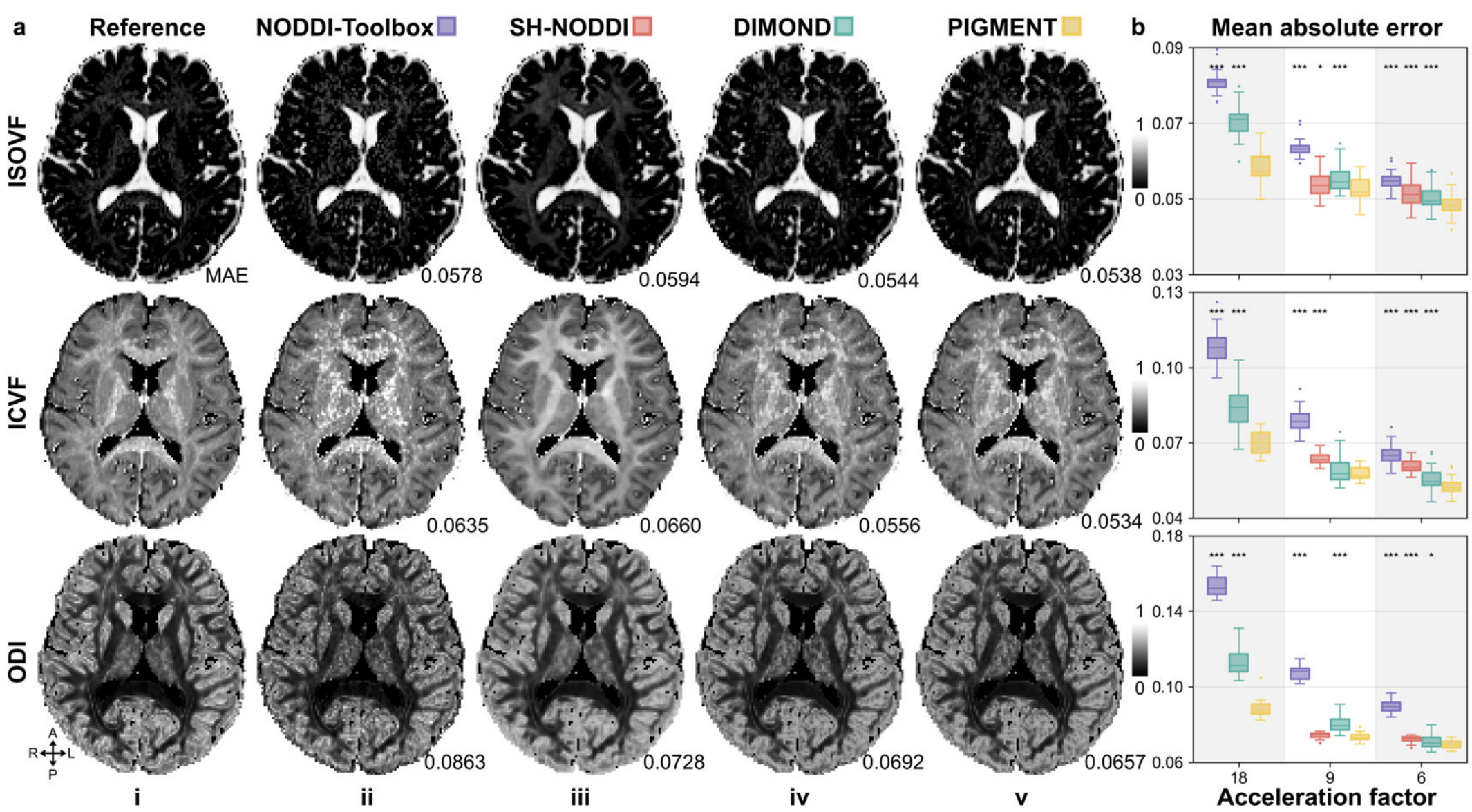

**Figure 4 | Robustness across diffusion encodings for neurite orientation dispersion and density imaging. a**, Exemplar axial maps of NODDI model metrics, including isotropic volume fraction (ISOVF), intracellular volume fraction (ICVF), and orientation dispersion index (ODI), are shown for a representative CHCP subject. Maps were generated using NODDI-Toolbox from the full multi-shell data of this CHCP subject (reference, i) and using other methods from accelerated data subsets with 6-fold acceleration (ii-v). **b**, Mean absolute errors (MAEs) of ISOVF, ICVF, and ODI relative to the reference are shown for 21 CHCP subjects across acceleration factors of 18, 9, and 6. Box plots illustrate the distribution of metric values, where the midline represents the median, box edges indicate the first and third quartiles, whiskers extend 1.5 times the interquartile range (IQR) with outliers plotted individually. Asterisks indicate the significance level of differences between each method and PIGMENT (two-sided paired *t*-test; $^{***}P < 0.001$, $^{**}P < 0.01$, $^{*}P < 0.05$).

PIGMENT consistently delivered high-quality estimations for advanced microstructural models (Figs. 3–4). For diffusion kurtosis imaging (DKI), where estimating 22 parameters from accelerated data typically poses a highly ill-conditioned inverse problem, conventional fitting (OLS) yielded maps severely corrupted by noise (Fig. 3a). In contrast, PIGMENT robustly recovered high-fidelity maps that closely resembled the reference across all acceleration levels. Quantitatively, PIGMENT achieved substantial performance gains, reducing errors by over 40% relative to OLS for DKI-derived metrics (e.g., 53% for FA, 71% for radial kurtosis (RK) at 6-fold acceleration), with significant improvement over competing methods (Fig. 3b). Similarly, for NODDI, PIGMENT consistently achieved the lowest MAE across all acceleration factors, significantly outperforming competing methods (Fig. 4b). Notably, PIGMENT demonstrated superior robustness at extreme acceleration factors (e.g., 18-fold with only 5 directions per shell), a regime where SH-based methods like SH-NODDI[31] fundamentally broke down due to insufficient angular sampling for spherical harmonic computation.

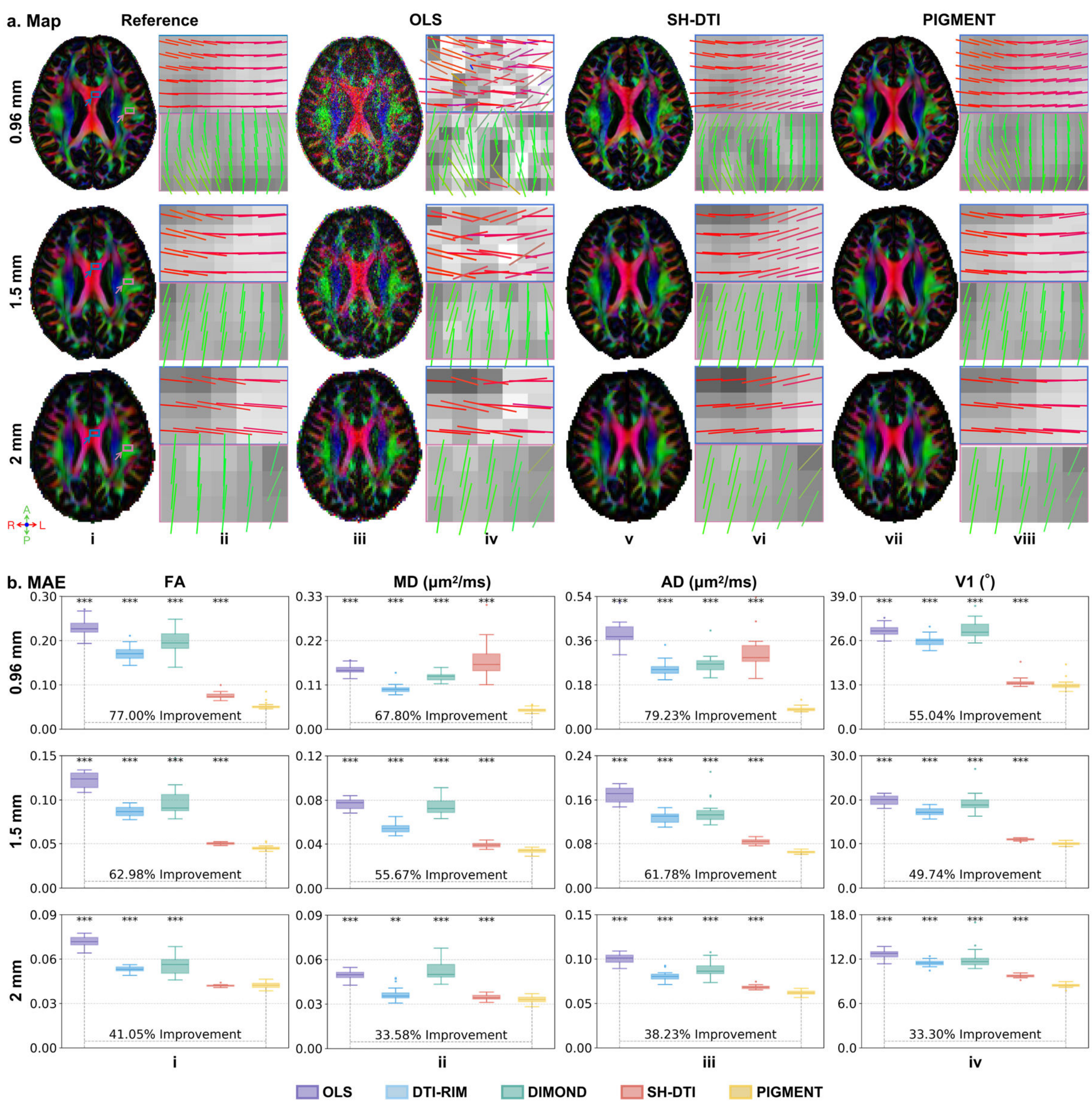

**Figure 5 | Robustness across spatial resolutions. a**, Exemplar axial maps of diffusion tensor model metric fractional anisotropy (FA) color-encoded by primary eigenvector (V1, red: left-right, green: anterior-posterior, blue: superior-inferior) along with V1 overlaid on FA of two enlarged regions of interest in the corpus callosum (blue boxes, left-right direction) and cingulum (pink boxes, anterior-posterior direction) are shown. Maps were generated using OLS from the full single-shell data of THU-3T dataset (reference, i) and using other methods from accelerated data subsets with 5-fold acceleration (ii-iv) at three empirically acquired spatial resolutions (0.96-, 1.5-, and 2-mm isotropic) of a representative subject from the THU-3T dataset. **b**, The mean absolute errors (MAEs) of FA (i), mean diffusivity (MD, ii), axial diffusivity (AD, iii), and V1 (iv) relative to the reference within the white matter mask, as well as their percentage reductions compared with baseline OLS MAEs, are shown across 20 subjects at three different spatial resolutions (0.96, 1.5 and 2 mm). Box plots illustrate the distribution of metric values, where the midline represents the median, box edges indicate the first and third quartiles, and whiskers extend 1.5 times the interquartile range (IQR) with outliers plotted individually. Asterisks indicate the significance level of differences between each method and PIGMENT (two-sided paired *t*-test; $^{***}P < 0.001$, $^{**}P < 0.01$, $^{*}P < 0.05$).

PIGMENT maintains robust generalizability and stable accuracy across diverse spatial resolutions. Using the THU-3T dataset, we evaluated PIGMENT on data acquired at 0.96-, 1.5-, and 2.0-mm isotropic resolutions. PIGMENT effectively mitigated the varying degrees of noise amplification inherent to different resolutions, demonstrating exceptional stability, particularly in challenging high-resolution regimes. Qualitatively, when processing the very sparse diffusion encoding of one b = 0 volume and six DWIs, conventional OLS yielded noisy and incoherent maps (Fig. 5a, ii), and SH-DTI exhibited substantial orientation bias (Fig. 5a, iii). In stark contrast, PIGMENT preserved anatomical details in key pathways like the corpus callosum and cingulum without introducing blurring (Fig. 5a, iv). Quantitatively, PIGMENT consistently achieved significantly lower MAEs across all tensor metrics ($P < 0.05$, two-sided paired *t*-test; Fig. 5b). While other methods deteriorated rapidly with increasing spatial resolution due to diminishing signal-to-noise ratio (SNR), PIGMENT maintained stable accuracy even at submillimeter scales. Notably, the relative performance advantage of PIGMENT over the OLS baseline scaled inversely with voxel size, widening from an average error reduction of 36.5% (across all four metrics) at 2.0-mm scale to a substantial 69.8% reduction at the highly demanding 0.96-mm scale.

## Preserving microstructural patterns across spatiotemporal dimensions

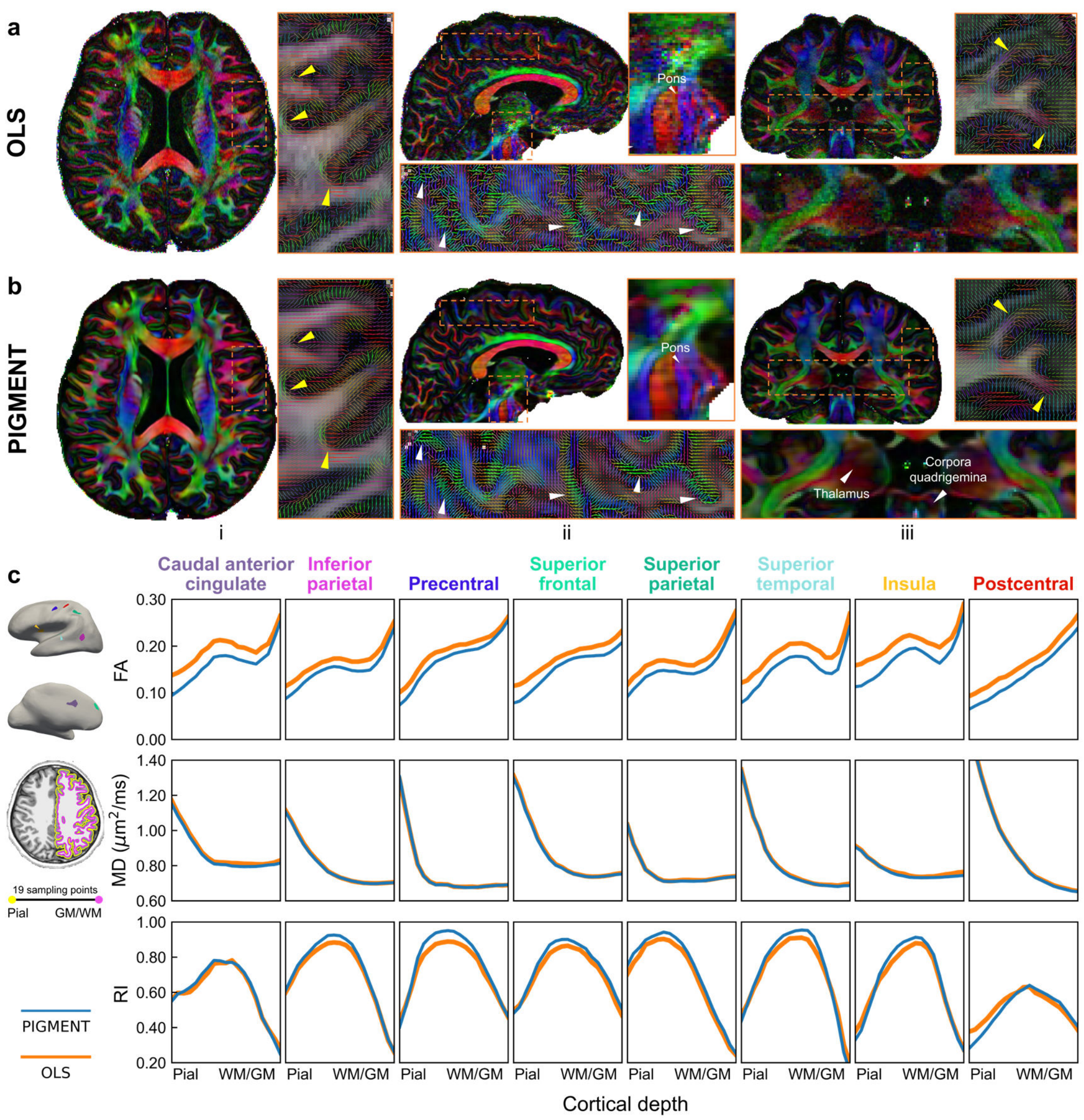

**Figure 6 | Performance on submillimeter data. a–b**, Exemplar axial (i), sagittal (ii) and coronal (iii) maps of fractional anisotropy (FA) color-encoded by primary eigenvector (V1, red: left-right, green: anterior-posterior, blue: superior-inferior), together with magnified views of selected regions (e.g., gyrus, gyral crowns, pons and corpora quadrigemina) are shown. Yellow arrows denote anatomical detail consistently resolved in both methods, whereas white arrows highlight improvements achieved by PIGMENT. Maps were generated from the full Oxford-Meso dataset using OLS (a) and PIGMENT (b). **c**, Profiles of FA, mean diffusivity (MD) and radiality index (RI) across 19 equidistant cortical depths, from the pial surface to the gray matter (GM)–white matter (WM) interface (middle left) are shown for eight representative ROIs (color-coded and displayed on the inflated cortical surfaces, upper left), comparing OLS (orange lines) and PIGMENT (blue lines).

PIGMENT is particularly effective in high-resolution regimes, where resolving microanatomical cortical features becomes challenging. We evaluated PIGMENT's biological utility for advanced neuroscience research under challenging acquisition regimes. Specifically, we investigated its capacity to resolve spatially structured cortical features from submillimeter data and to track temporally evolving neurodevelopmental trajectories using highly accelerated protocols requiring only six DWIs. PIGMENT yields exceptional visual clarity for *in vivo* microarchitectures. On the 0.7-mm isotropic diffusion data from the Oxford-Meso dataset, PIGMENT reconstructed white-matter microstructure with high fidelity and minimal blurring, clearly resolving anatomical features typically discernible only in high-resolution acquisitions, such as U-shaped short association fibers (Fig. 6a–b, i, yellow arrows) and fiber fanning at gyral crowns (Fig. 6a–b, iii, right, yellow arrows). Furthermore, PIGMENT enhanced the microstructural delineation of both deep brain structures and the cortex with superior noise suppression, more clearly delineating the fiber trajectories within the pons (Fig. 6a–b, ii, right, white arrows), thalamus and corpora quadrigemina (Fig. 6a–b, iii, left, white arrows), as well as yielding more coherent fiber orientations in the cortex (Fig. 6a–b, ii, left, white arrows).

This structural fidelity facilitates the precise quantification of laminar cortical microstructural features as biomarkers. We examined the profiles of FA, MD, and radiality index (RI) across cortical depths in eight representative ROIs (Fig. 6c). The PIGMENT-derived profiles closely aligned with the OLS baseline and prior literature[21,38], showing increasing FA and decreasing MD from the pial surface to the GM-WM interface, and reproducing key *in vivo* physiological signatures such as the RI peak at intermediate depths and region-specific FA extrema (e.g., in the insula)[21]. Moreover, the excellent agreement of MD and RI profiles between methods demonstrates PIGMENT's unbiased microstructural estimation. Notably, in intrinsically low-anisotropy regions (e.g., near the pial surface), PIGMENT yielded lower FA values compared to standard OLS. Since the noise can spuriously inflate apparent FA in such areas, this reduction indicates that PIGMENT effectively restores the true structural anisotropy.

Extending beyond static high-resolution imaging, PIGMENT faithfully preserves developmental microstructural trajectories from highly accelerated pediatric data. We applied PIGMENT to a 10-fold accelerated pediatric dataset ($n$=70, aged 0–5 years) and characterized age-dependent white matter maturation using ROI- and tract-based analyses (Fig. S9). PIGMENT reliably preserved age-dependent dynamics despite a substantial distribution shift relative to the training data, which strictly excluded subjects aged 0–5 years, thereby accurately characterizing developmental states throughout early childhood. Visual inspection confirmed robust reconstruction and consistent morphology of key structures (e.g., thalamocortical tract) across all age groups (Fig. 7a–c), aligning with the consensus that major white matter pathways are established prenatally. Biologically, PIGMENT-derived metrics captured expected developmental patterns, including age-related FA increases, RD decreases, and characteristic biphasic trajectories, mirroring the established literature on white matter maturation (Fig. 7d). Comparative analysis revealed that the thalamocortical tract exhibited higher FA, lower RD, and a shorter rapid growth phase than the frontal lobe, consistent with the functional hierarchy where sensorimotor pathways mature earlier than higher-order regions. Similarly, deep white matter (DWM) displayed higher FA, lower RD, and faster maturation rates than superficial white matter (SWM), aligning with known centrifugal developmental gradients[39]. Quantitatively, PIGMENT-derived trajectories were nearly identical to the reference, with a CCC exceeding 0.94, demonstrating remarkable consistency with fully acquired data at this 10-fold acceleration. These results unlock the ability to track temporally evolving neurodevelopmental trajectories from ultra-fast acquisitions and support the clinical prospect of rapid, sedation-free *in vivo* imaging for children.

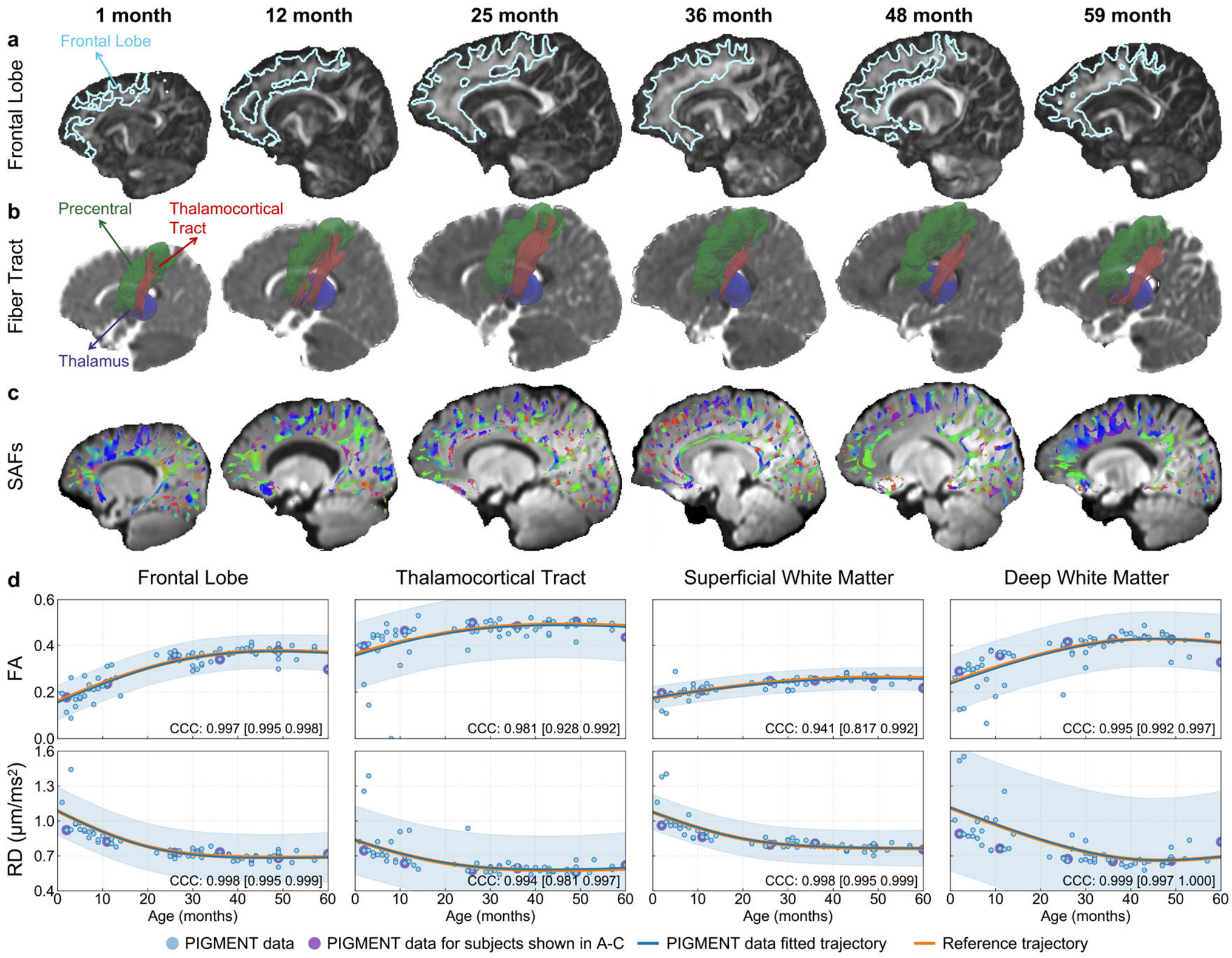

**Figure 7 | Performance on pediatric data. a–c,** Exemplar sagittal maps of diffusion tensor model metrics, including fractional anisotropy (FA) and radial diffusivity (RD), and the mean diffusion-weighted image (DWI), are shown for representative subjects aged 1, 12, 25, 36, 48 and 59 months (purple dots in d). **d,** Developmental trajectories of mean FA and RD within the frontal lobe (outlines in a), the thalamocortical tract (red in b) that passes through both the thalamus (blue in b) and precentral gyrus (green in b), the superficial white matter (partially represented by short-range association fibers, SAFs, in c), and the deep white matter are shown as a function of age for the reference (orange curves) and PIGMENT on sub-sampled data (purple and blue dots, blue curves). Each subplot reports the concordance correlation coefficient (CCC) between PIGMENT-derived results and the reference, with the 95% confidence interval (CI).

## Bridging the translational gap between quantitative microstructure biomarkers and clinical practice

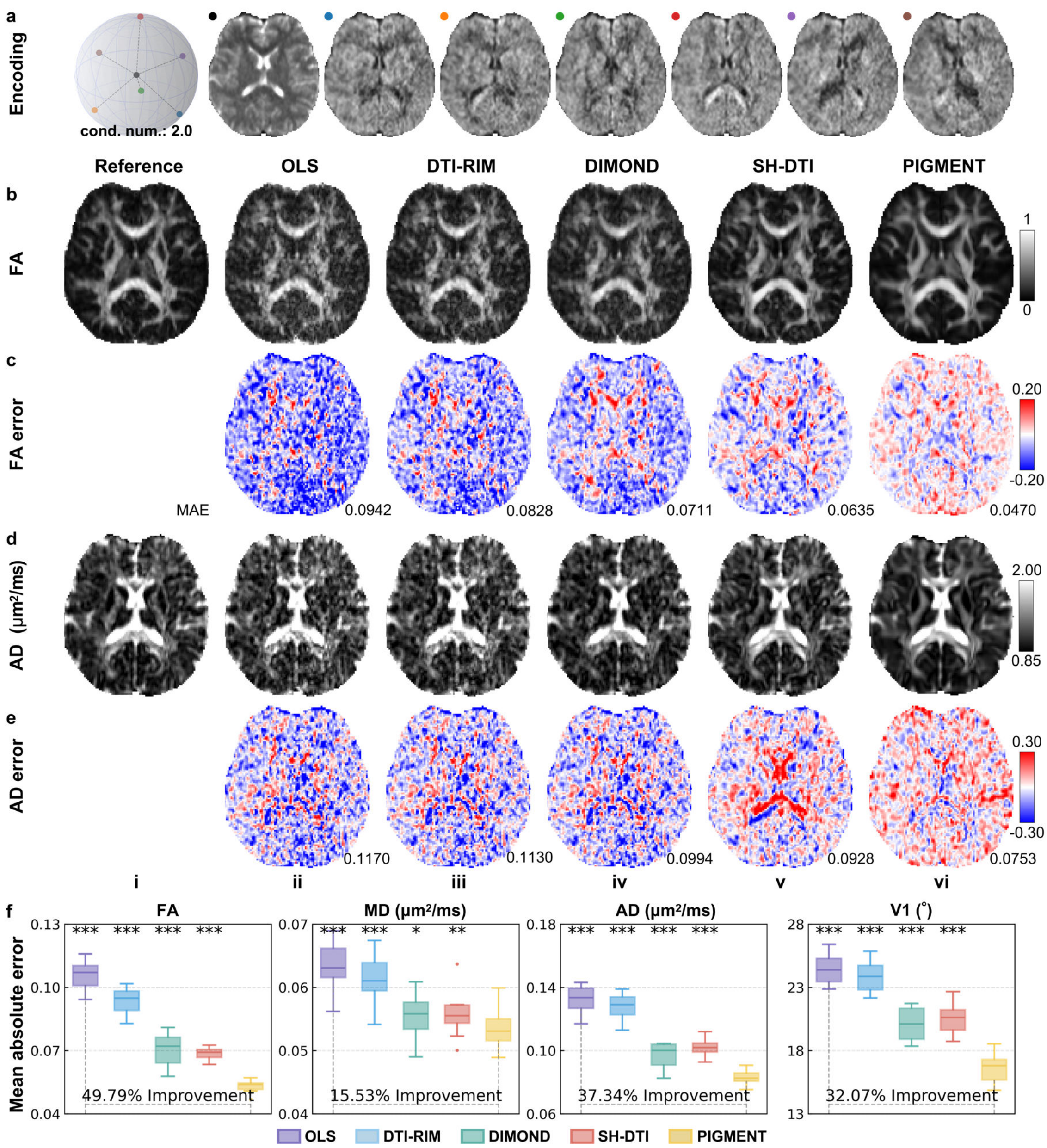

**Figure 8 | Performance on 0.55-Tesla data. a**, The empirical diffusion encoding scheme used by the 0.55-Tesla MAGNETOM Free.Max scanner includes a single b = 0 volume (black dots) and six diffusion-weighted volumes (color dots). **b–e**, Exemplar axial maps of diffusion tensor model metrics fractional anisotropy (FA, b) and axial diffusivity (AD, d), along with corresponding difference maps relative to the reference (c: FA, e: AD, numbers denote the mean absolute error (MAE) within the white matter mask), are shown for a representative subject. Maps were generated using OLS from the full single-shell data (i, reference) and using other methods from accelerated data subsets with 12-fold acceleration (ii-vi). **f**, MAEs of FA, mean diffusivity (MD), AD, and primary eigenvector (V1) relative to the reference, and their percentage reductions compared with baseline OLS MAEs, are shown for 10 subjects. Box plots illustrate the distribution of metric values, where the midline represents the median, box edges indicate the first and third quartiles, whiskers extend 1.5 times the interquartile range (IQR) with outliers plotted individually. Asterisks indicate the significance level of differences between each method and PIGMENT (two-sided paired *t*-test; $^{***}P < 0.001$, $^{**}P < 0.01$, $^{*}P < 0.05$).

PIGMENT enables quantitative diffusion MRI under clinically constrained acquisition conditions. We evaluated PIGMENT on 0.55-Tesla low-field MRI, an emerging platform poised to expand access to neuroimaging in medically underserved regions, yet fundamentally limited by stringent hardware constraints. To minimize concomitant fields, the six-direction encoding scheme in low-field scanners is heavily constrained, yielding a high condition number of 2.0 (Fig. 8a) that poses mathematical challenges for stable tensor estimation. Despite these sub-optimal acquisition conditions and the intrinsically low SNR of low-field systems, PIGMENT recovered tensor model metrics with substantially higher fidelity than competing methods. Qualitatively, the PIGMENT-derived FA and MD maps preserved fine structural details and exhibited smaller residuals without apparent systematic bias (Fig. 8b–e). In a representative subject, PIGMENT reduced the mean absolute error (MAE) of FA by 25.98% compared to the second-best method (SH-DTI) and by 50.11% compared to standard OLS. Similarly, for AD, it achieved MAE reductions of 18.86% and 35.64% relative to SH-DTI and OLS, respectively. Furthermore, at the cohort level, PIGMENT achieved significantly lower MAEs than all comparison methods across all tensor model metrics ($P < 0.05$, two-sided paired t-test; Fig. 8f). Specifically, compared to conventional OLS, PIGMENT yielded significant error reductions of 49.79%, 15.53%, 37.34%, and 32.07% for FA, MD, AD, and V1, respectively. Ultimately, these results demonstrate PIGMENT's stable and accurate microstructural quantification capabilities on low-field scanners, providing the high-fidelity microstructural maps requisite for deriving downstream biomarkers in resource-limited settings.

Under severe scan-time constraints, we assessed PIGMENT on ultra-sparse clinical screening protocols. Conventional DTI fundamentally requires at least six non-collinear directions. However, by leveraging a learned generative prior, PIGMENT successfully estimated tensors from just one to five DWIs on the CHCP dataset (Fig. 9a). These highly accelerated PIGMENT estimates achieved lower mean absolute errors than the standard OLS utilizing all six directions, with accuracy improving monotonically as the number of DWIs increased (Fig. 9b). Crucially, even at an extreme sparsity of three DWIs, PIGMENT supported reliable downstream tractography. The resulting structural connectivity matrices exhibited excellent agreement with fully sampled references (CCC = 0.973, $R = 0.974$; Fig. 9c), demonstrating PIGMENT's viability for network-level analyses.

PIGMENT preserves clinically relevant biomarkers and tractography in highly accelerated neuro-oncology protocols. Using a six-direction protocol, PIGMENT generated high-fidelity FA maps that sharply delineated the boundaries of a WHO Grade I meningioma while preserving intricate white matter architecture without systematic bias (Fig. 10a–b, vi). In contrast, results from conventional OLS (Fig. 10a, ii) and DTI-RIM[40] (Fig. 10a, v) were severely compromised by noise, obscuring critical anatomical details. Quantitatively, across 25 patients, PIGMENT-derived mean tensor metrics (FA, MD, and AD) within the tumor region closely matched the reference (Fig. 10d, vi), whereas other baseline methods exhibited substantial deviations. Notably, even when the encoding directions were reduced to three, PIGMENT maintained coherent tumor contrast and outperformed the six-direction OLS baseline in biomarker consistency (higher CCC values). Supported by consistent findings in the independent WCH-Tumor dataset (Fig. 9d), PIGMENT's faithful reproduction of these pathological biomarkers[41] holds significant promise for facilitating rapid yet precise clinical workflows, such as tumor grading[42] and prognostic assessment[43].

Beyond microstructural mapping, PIGMENT demonstrated significant translational potential in delineating structural connectivity in ultra-fast scanning, which is critical for neurosurgical planning and diagnostic characterization. Tractography of the tumor-affected superior longitudinal fasciculus (SLF) derived from PIGMENT exhibited high anatomical fidelity compared to the reference (Fig. 10c), effectively reconstructing the complete fiber trajectory despite peritumoral edema and infiltration. Notably, even when the number of encoding directions was reduced to three, PIGMENT successfully resolved the principal fiber trajectories adjacent to the tumor. This capability underscores PIGMENT's robustness in extremely low-direction clinical settings, suggesting it can support the delineation of eloquent pathways even under highly constrained acquisition protocols[44].

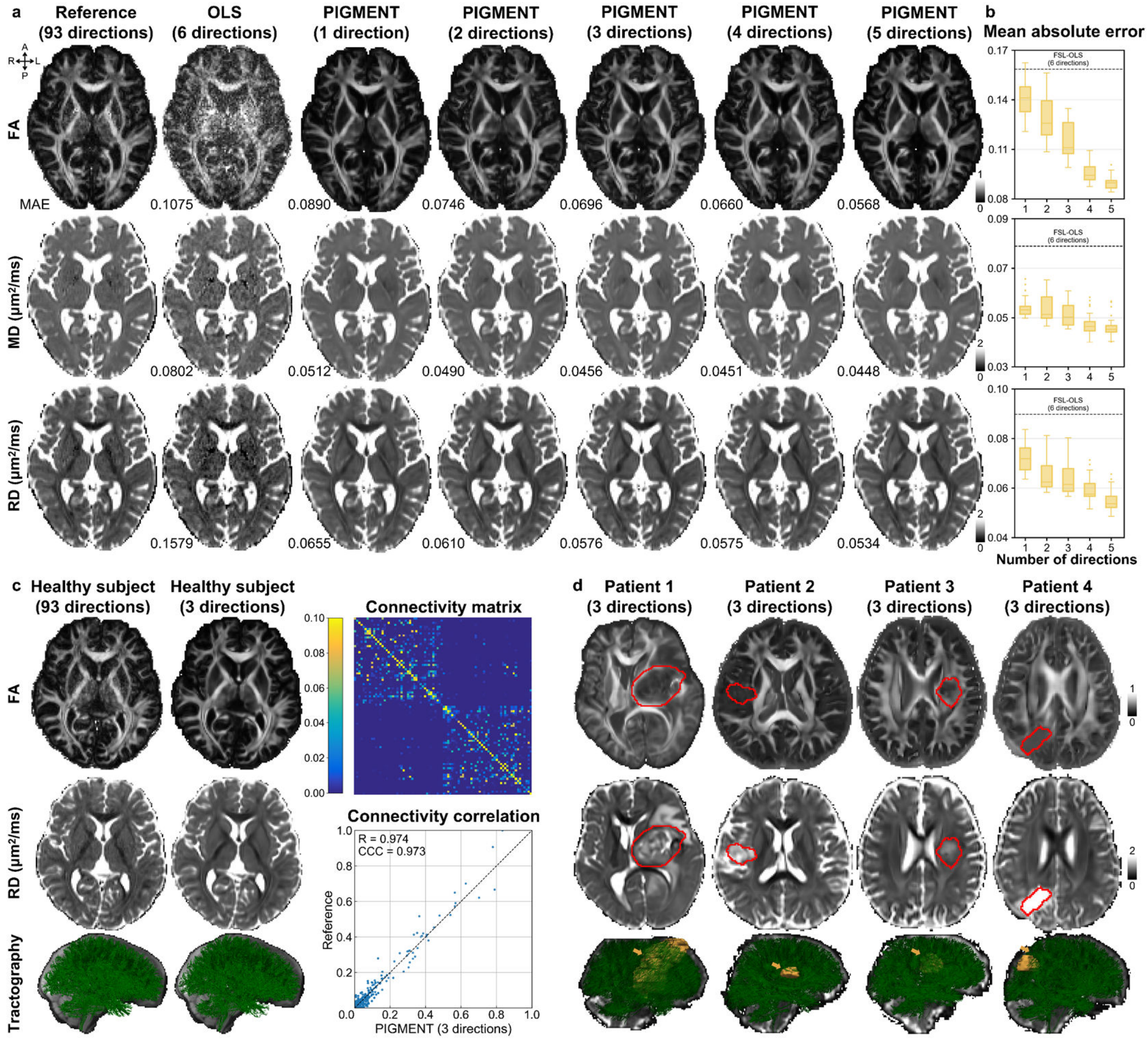

**Figure 9 | Microstructure estimation from ultra-sparse clinical acquisition. a**, Exemplar axial maps of diffusion tensor model metrics fractional anisotropy (FA), mean diffusivity (MD), and radial diffusivity (RD) are shown for a representative CHCP subject. Maps were generated using OLS on the full single-shell data (reference), OLS on six diffusion-weighted images (DWIs), and PIGMENT on one to five DWIs (numbers denote the mean absolute error (MAE) within the white matter mask). Note that six DWIs represent the minimum requirement for tensor modeling. **b**, MAEs of FA, MD, and RD relative to the reference are shown for 21 CHCP subjects from one to five DWIs using PIGMENT. Box plots illustrate the distribution of metric values, where the midline represents the median, box edges indicate the first and third quartiles, and whiskers extend 1.5 times the interquartile range (IQR) with outliers plotted individually. **c**, Exemplar axial FA and RD maps and whole-brain tractography are shown for a representative healthy subject. Maps were generated using OLS on full single-shell data for the healthy subject (reference), and PIGMENT on clinical DWIs along three orthogonal directions. The structural connectivity matrix derived from three-direction PIGMENT data is shown, and its connection-wise correlation with the reference is quantified by Pearson's $r$ and the concordance correlation coefficient (CCC). **d**, Exemplar axial maps of FA, RD and whole-brain tractography are shown for four patients with brain tumors. Maps were generated via PIGMENT using clinical three-direction DWIs. Tumors are indicated in red on the FA and RD maps, and in orange on the tractography.

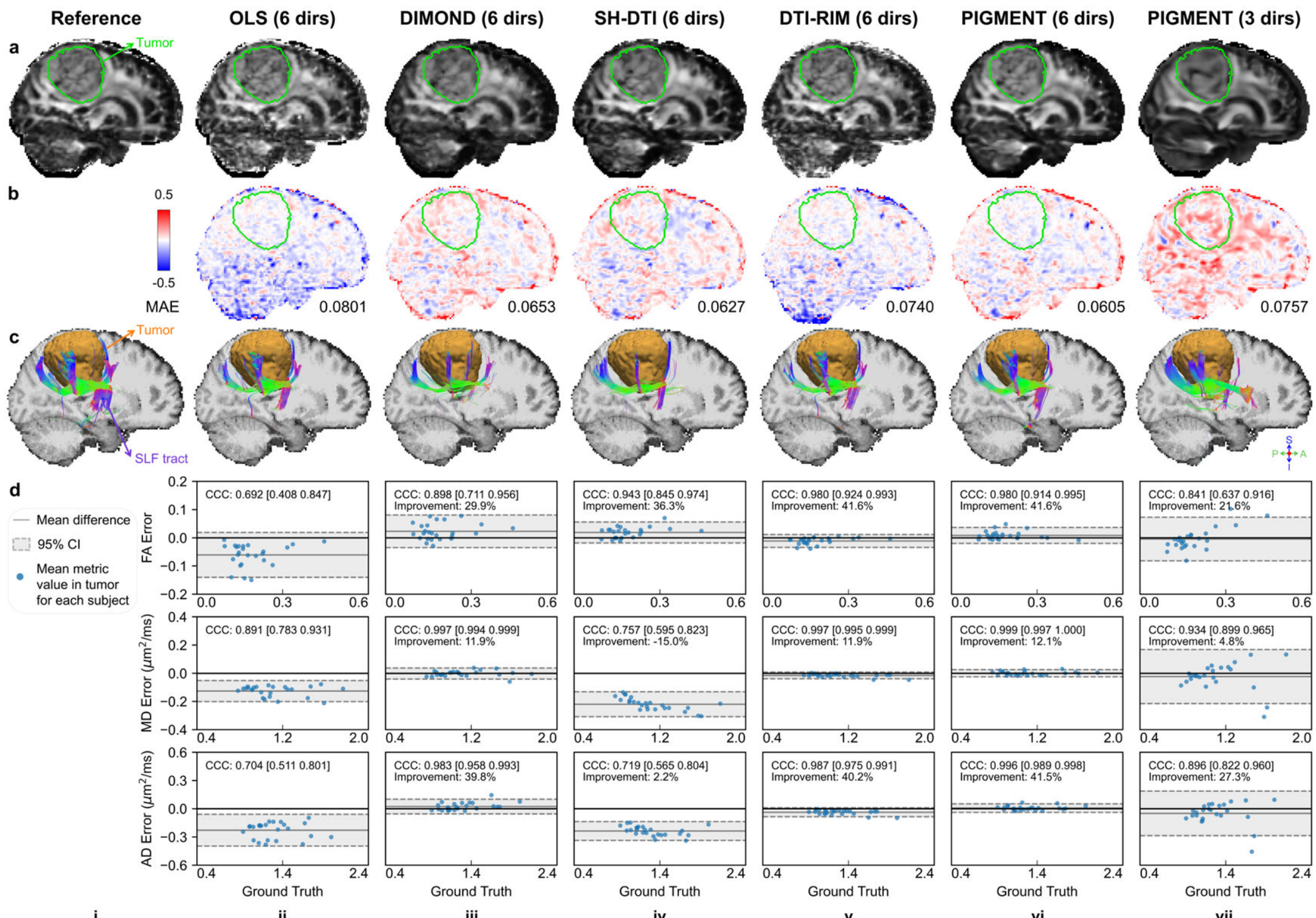

**Figure 10 | Performance on neuro-oncologic data. a–c**, Exemplar sagittal maps of diffusion tensor model metric fractional anisotropy (FA, a), FA difference maps relative to the reference (b, numbers denote the mean absolute error (MAE) within normal brain tissue), as well as the right superior longitudinal fasciculus (SLF) overlaid on the T1w image (c), are shown for a representative subject from the Ghent-Tumor dataset, with the tumor indicated by green (a, b) and orange (c). The maps and fiber tracts were generated using OLS from the full single-shell data (reference, i), using other methods from 5-fold accelerated data subsets (6 directions, ii-vi) and using PIGMENT from a 10-fold accelerated subset (3 directions, vii). **d**, The errors and concordance correlation coefficients (CCC) of mean FA, mean diffusivity (MD), and axial diffusivity (AD) within the tumor mask relative to the reference, as well as the percentage improvement in CCCs compared with baseline OLS, are shown across 25 subjects. Bland–Altman plots illustrate the mean bias (solid gray lines) and 95% confidence interval (CI) for the errors (dashed gray lines).

## DISCUSSION

In this study, we introduce PIGMENT, a physics-informed foundation model that generalizes across heterogeneous diffusion MRI datasets, thereby resolving the fundamental trade-off between acquisition feasibility and estimation fidelity. Algorithmically, PIGMENT achieves this by learning a population-scale latent generative prior over microstructural parameter maps, which is rigorously constrained at test time by each subject's measured diffusion signals via explicit forward models. Comprehensive evaluations across disparate imaging protocols and microstructural models demonstrate that PIGMENT consistently yields high-fidelity microstructural maps. Building upon this methodological generalizability, PIGMENT enables profound spatiotemporal insights into brain microarchitecture using highly accelerated, pediatric-friendly acquisitions. Ultimately, by facilitating robust tensor imaging and the derivation of diagnostic tumor biomarkers in resource-constrained environments, PIGMENT establishes a translational framework for democratizing precision neuroimaging, and is therefore poised to unlock the latent scientific value of historically suboptimal diffusion data.

The deployment of foundation models to serve as generative priors represents a paradigm shift in brain microstructure modeling. While conventional numerical optimization is mathematically optimal under ideal conditions, its results are substantially corrupted by noise (Fig. 3a, FA) and prone to generating stochastic outliers (Fig. 3a, AK, RK). Although end-to-end deep learning approaches (e.g., SH-DTI) can mitigate noise, they struggle to generalize to unseen data, particularly for pathological cases, which are inherently rare and subject-specific. For instance, while such approaches achieved lower MAE globally (Fig. 10b), they failed to preserve the heterogeneity of lesion biomarkers (Fig. 10d). This discrepancy between pixel-level metrics and clinical fidelity represents a pervasive challenge in medical artificial intelligence, and highlights the vulnerability of pure end-to-end models to severe over-smoothing when encountering out-of-distribution data. Consequently, synergizing prior information with data consistency to ensure biophysical fidelity and prevent generative hallucinations serves as the core design philosophy of PIGMENT. We also demonstrate that PIGMENT's data-driven generative prior achieves superior regularization compared to heuristic manual designs (e.g., DESIGNER-CWLS) and learned iterative solvers (e.g., DTI-RIM). Crucially, this generative design solidifies its methodological generalizability, ensuring high reconstruction fidelity regardless of diverse diffusion encodings or specific acquisition protocols, yielding biologically plausible (Fig. S8) and clinically meaningful results (Fig. 9).

These methodological advances endow PIGMENT with the transformative capability to recover quantitative microstructural metrics from highly constrained clinical acquisitions. Conventional algorithms routinely fail under such mathematically underdetermined conditions, creating a severe analytical bottleneck. Consequently, routine scans are relegated exclusively to qualitative visual inspection, leaving their rich microstructural information entirely unexploited[45,46]. Historically, reliable microstructural mapping required high-field systems and dense angular sampling, confining this technology to specialized research centers. In routine practice, such acquisitions remain impractical due to infrastructural limitations and strict time constraints. Specifically, cost-efficient low-field scanners suffer from intrinsic signal starvation and ill-conditioned gradient tables resulting from the mitigation of concomitant magnetic fields. Similarly, ubiquitous three-direction screening protocols lack the angular density required for standard mathematical inversion. By leveraging a generative prior encoding broad human neuroanatomy, PIGMENT acts as a powerful computational equalizer that resolves these physical and mathematical barriers. It yields unbiased neuro-oncological biomarkers and delineates eloquent white matter pathways for surgical planning directly from ultra-fast (Figs. 9–10) or low-field scans (Fig. 8)[47]. This paradigm shift democratizes precision diagnostics for community hospitals without requiring prohibitive hardware upgrades. More profoundly, it unlocks the immense scientific potential of vast historical clinical archives, empowering population-scale neuroimaging investigations from legacy datasets previously constrained by an insufficient number of diffusion directions for robust quantitative analysis.

The ability to derive comprehensive tensor and multi-compartment metrics from ultra-sparse protocols (e.g., three-direction DWIs) fundamentally expands the operational limits of acute and interventional neuroimaging. In time-critical scenarios, such as acute ischemic stroke triage or intraoperative neurosurgical navigation[48], lengthy high-angular-resolution acquisitions are strictly prohibitive. Consequently, clinicians rely on macroscopic and qualitative imaging modalities, often missing subtle microstructural penumbras or structurally displaced eloquent pathways distorted by peritumoral edema[49,50]. By extracting high-fidelity quantitative biomarkers from sequences acquired in merely minutes (Figs. 9–10), PIGMENT seamlessly converts routine anatomical triage into a rapid, quantitative precision-medicine tool. This empowers neurosurgeons with microscopically accurate maps of fiber integrity and tumor infiltration margins directly within the operating theater, delineating maximal safe resection boundaries and profoundly improving postoperative neurological outcomes in time-sensitive clinical environments.

In pediatric neuroimaging, PIGMENT substantially accelerates microstructural mapping, mitigating the risks associated with sedation and ensuring high patient compliance without compromising quantitative accuracy. For instance, despite this drastic acceleration, PIGMENT successfully replicated established developmental trajectories in the frontal lobe and thalamocortical tracts (Fig. 7)[51,52]. Crucially, we extended these quantitative insights to characterize the superficial white matter (SWM)[53], a region notoriously difficult to map without prohibitively lengthy high-angular-resolution acquisitions[54]. To our knowledge, PIGMENT provides the first robust mapping of SWM maturation in early childhood (0–5 years), a cohort that conventionally requires sedation for high-fidelity diffusion imaging. This capability transforms historically lengthy research sequences into viable screening tools, facilitating the early, accessible detection of neurodevelopmental disorders such as autism[55]. Furthermore, PIGMENT functions as a seamless, “plug-and-play” engine. It synthesizes full-q-space diffusion MRI outputs natively compatible with ubiquitous analysis pipelines (e.g., FSL[56], MRtrix3[57], and FreeSurfer[58]), streamlining downstream tractography without disrupting established clinical workflows. Ultimately, by democratizing precision diagnostics and unlocking massive archives of previously unquantifiable clinical data, PIGMENT bridges the critical gap between fundamental neuroscientific discovery and routine clinical translation.

The strong performance of PIGMENT in few-shot multimodal synthesis demonstrates that the pre-trained network has learned a robust and transferrable structural prior from diffusion MRI data, which inherently contains rich and complementary image contrasts. Specifically, non-diffusion-weighted images naturally show T2w contrast, whereas DWIs exhibit strong tissue-contrast characteristics similar to T1w images, with additional water-suppression effects resulting in a FLAIR-like contrast. Our experiments confirm this relationship, showing that DWIs are generated with the highest quality, followed by T2w and then T1w images (Fig. S5). PIGMENT effectively compresses this rich and heterogeneous imaging information into a compact and highly informative feature space, capturing underlying biophysical and contrast-related properties, thereby enabling strong cross-modal generative capability and supporting a wide range of downstream tasks.

Beyond individual analysis, this unified and expressive latent space enables PIGMENT to drive a paradigm shift in multi-center neuroimaging. Historically, the immense hardware heterogeneity across clinical sites, spanning varying field strengths, gradient systems, and vendor-specific artifacts, has posed an intractable harmonization challenge, severely confounding population-level statistical analyses[59]. Based on the robustness across multi-center and multi-protocol validation, PIGMENT can effectively decouple microstructural quantification from hardware variance by projecting diverse, unharmonized diffusion measurements into a universal, biologically grounded latent feature space. This scanner-agnostic capability positions PIGMENT as a universal translational framework, enabling researchers to pool fragmented, multi-site datasets into cohesive, ultra-large-scale cohorts with unprecedented statistical power. Consequently, it paves the way for establishing harmonized normative baselines for individual diagnosis, seamlessly bridging the translational gap between state-of-the-art research centers and routine clinical practice in community hospitals.

Several limitations warrant future investigation. First, the lack of a dedicated 0.55-Tesla patient cohort limits the direct validation of low-field diagnostic performance, necessitating future evaluations in real-world clinical populations. Second, the heterogeneous parameter spaces of various microstructural models currently require training separate frameworks for specific targets. Future work aims to align these into a shared latent space to build a universal, multi-compartment foundation model. Lastly, rather than relying solely on microstructural metrics for diagnosis, future work will directly exploit PIGMENT's intermediate latent representations to enable powerful, end-to-end disease prediction.

# METHODS

## Generative microstructure prior learning

PIGMENT learns a comprehensive prior of human brain microstructure by integrating a microstructure autoencoder and a generative denoising U-Net. To capture the complex 3D spatial coherence of microstructures, PIGMENT adopts a volumetric modeling approach. However, directly processing high-dimensional parametric volumes is computationally expensive. Thus, we propose to employ a 3D vector-quantized variational autoencoder consisting of an encoder $\mathcal{E}$ and a decoder $\mathcal{D}$ to map the high-dimensional parameter space into a compact, discrete latent space (Fig. 1a). Formally, the encoder maps the input volume to a latent space $z = \mathcal{E}(x)$ and the decoder reconstructs the parameters $\hat{x} = \mathcal{D}(z)$, where $z$ is quantized into discrete codes before decoding to ensure stable generative performance. The network parameters are jointly optimized by minimizing a composite objective function comprising reconstruction and quantization terms:

$$\mathcal{L}_{vq} = \mathcal{L}_{rec} + \mathcal{L}_{quant}. \tag{1}$$

To account for the distinct geometric properties of microstructure model parameters, we propose a new reconstruction loss $\mathcal{L}_{rec}$ that contains specific objectives for scalar intensities ($x_{scalar}$) and angular orientations ($x_{angular}$):

$$\mathcal{L}_{rec} = \mathcal{L}_{scalar} + \mathcal{L}_{angular}, \tag{2}$$

$$\mathcal{L}_{scalar} = \mathbb{E}[\|x_{scalar} - \hat{x}_{scalar}\|_1], \tag{3}$$

$$\mathcal{L}_{angular} = \mathbb{E}\left[1 - \langle \frac{x_{angular}}{\| x_{angular} \|_2}, \frac{\hat{x}_{angular}}{\| \hat{x}_{angular} \|_2} \rangle^2 \right]. \tag{4}$$

The standard quantization objective is employed to optimize the discrete codebook, thereby regularizing the learned representations against a fixed, meaningful set of discrete codes across datasets[60]:

$$\mathcal{L}_{quant} = \| \mathrm{sg}[z_e(x)] - e \|_2^2 + \beta \| z_e(x) - \mathrm{sg}[e] \|_2^2, \tag{5}$$

where $z_e(x)$ denotes the output of the encoder, $e$ represents the learnable codebook embeddings, and sg[·]denotes the stop-gradient operator. The hyperparameter $\beta$ regulates the commitment of the encoder to the codebook embeddings and was empirically set to 0.25.

Following spatial compression, we propose to train a latent denoising diffusion model to capture the underlying distribution of microstructural parameters within the compressed latent space. The diffusion model employs an attention-based 3D denoising U-Net, denoted as $\epsilon_\theta$, to predict the noise component injected into the latent representation at each time step ( $t = 1, \dots, T, T = 1000$ ). The network parameters $\theta$ are optimized via the standard denoising objective[61,62]:

$$\mathcal{L}_{diffusion} = \mathbb{E}_{\mathcal{E}(x), \epsilon, t}[\|\epsilon_\theta(z_t, t, S_0) - \epsilon\|_2^2], \tag{6}$$

where $z_t$ represents the noisy latent at timestep $t$, and $\epsilon \sim \mathcal{N}(0, \mathbf{I})$ is the Gaussian noise added to the signal. Crucially, the generative process is conditioned on the non-diffusion-weighted signal ($S_0$). Because the $S_0$volume is universally acquired across all diffusion MRI protocols and inherently features a high SNR, it provides robust anatomical contrast to guide the high-fidelity synthesis of microstructural maps.

## Physics-informed inference

PIGMENT generates the microstructure estimation of each individual by denoising and refining the microstructure latent using the pre-trained latent diffusion model and dual-constraint optimization, respectively. The generation procedure initiates from a random Gaussian noise $z_T \sim \mathcal{N}(0, \mathbf{I})$ and progressively recovers underlying anatomical structure using the denoising diffusion implicit model sampling trajectory[63]. Through this trajectory, noise is systematically removed by the network $\epsilon_\theta$, transforming the latent into a biophysically plausible representation:

$$z'_{t-1} = \sqrt{\bar{\alpha}_{t-1}}\hat{z}_0(z_t) + \sqrt{1 - \bar{\alpha}_{t-1} - \delta_t^2}\, \epsilon_\theta(z_t, t, S_0) + \delta_t \epsilon, \tag{7}$$

where the estimation of the noise-free latent is given by:

$$\hat{z}_0(z_t) \simeq \mathbb{E}[z_0|z_t] = \frac{z_t - \sqrt{1 - \bar{\alpha}_t}\epsilon_\theta(z_t, t, S_0)}{\sqrt{\bar{\alpha}_t}}. \tag{8}$$

Here $\bar{\alpha}_t$ denotes the cumulative noise schedule coefficient, and $\delta_t$ controls the stochasticity of the sampling process.

The above process generates plausible samples from the prior, but requires further constraints to ensure these samples accurately reflect the individual acquisition data. Therefore, we propose a dual-domain optimization strategy that refines the latent estimate by enforcing both physical and manifold consistency integrated within sampling steps. Specifically, to enforce consistency between the estimation and the acquired diffusion signals $\boldsymbol{S}$, we introduce a physics-informed data consistency constraint, which optimizes the difference between the acquired signals and the synthetic signals derived from the forward microstructural model based on the parameter estimated at the current timestep:

$$\mathcal{L}_{consistency} = \mathbb{E}_{patches}\left[\left\|\boldsymbol{S} - Model\left(\mathcal{D}\big(\hat{z}_0(z_t)\big)\right)\right\|_2^2\right]. \tag{9}$$

However, optimizing Eq. (9) solely based on signal fidelity risks overfitting to acquisition noise or drifting off the data manifold. To ensure the solution remains within the learned prior distribution, a denoising score matching consistency objective[64] is imposed on the latent space, which enforces that when the optimized parameter $\hat{x}_{opt} = \mathcal{D}\big(\hat{z}_0(z_t)\big)$ is re-encoded via the encoder $\mathcal{E}$ and perturbed with noise $n_k$, the diffusion model can still accurately predict the added noise:

$$\mathcal{L}_{manifold} = \mathbb{E}_{k,n_k}\left[\left\|\epsilon_\theta\big(\mathcal{E}(\hat{x}_{opt}) + n_k, k\big) - \epsilon\right\|_2^2\right], \tag{10}$$

where $n_k$ is Gaussian noise added at a randomly selected timestep $k \in [0,100]$. Since direct optimization of the diffusion objective involves computationally expensive backpropagation through the U-Net Jacobian, we employ the gradient approximation proposed in [64]:

$$\nabla_{\mathbf{z}}\mathcal{L}_{\text{manifold}} \approx \mathbb{E}_{k,\epsilon}\left[w(k)\Big(\epsilon_\theta\big(\mathcal{E}(\hat{x}_{opt}), k\big) - \epsilon\Big)\right], \tag{11}$$

where $w(k) = 1 - \bar{\alpha}_k$ is a time-dependent weighting factor that emphasizes alignment at higher noise levels, and gradients are not propagated through $\epsilon_\theta$. This dual-domain optimization strategy ensures both signal fidelity in the physical image domain and manifold consistency in the latent space:

$$\mathcal{L}_{dual} = \mathcal{L}_{consistency} + \lambda\mathcal{L}_{manifold}. \tag{12}$$

Finally, the stochastic resampling technique[65] is utilized to fuse $z'_t$, the original latent representation at timestep *t*, with the latent generated by encoding the optimized parameter $\hat{x}_{opt}$:

$$z_{t-1} = (1-\gamma)z'_{t-1} + \gamma\left(\sqrt{\bar{\alpha}_{t-1}}\mathcal{E}(\hat{x}_{opt}) + \sqrt{1-\bar{\alpha}_{t-1}}\epsilon\right), \quad (13)$$

where $\gamma$ acts as a weighting factor balancing the prior preservation and data consistency.

**Training datasets and pre-processing**

The training cohort comprises data from five publicly available datasets and two in-house datasets. Key acquisition parameters, including spatial resolution, b-values, and diffusion encoding directions for each dataset are summarized in Table S1. For the publicly available datasets, we utilize the officially pre-processed data from the Alzheimer's Disease Neuroimaging Initiative (ADNI) database[66-68], the Parkinson's Progression Markers Initiative (PPMI) database[69,70], the Southwest University Longitudinal Imaging Multimodal (SLIM) brain data repository[71], and both the distinct 3T and 7T datasets from the human connectome project (HCP) WU-Minn-Ox Consortium[72]. For the two in-house datasets, we provide the imaging protocols and pre-processing pipeline as follows.

***THU-5T dataset***

In accordance with local ethics, 2-mm isotropic diffusion MRI data were acquired from 111 subjects using a 2D echo-planar imaging (EPI) sequence on a uMR Jupiter 5.0-Tesla scanner (United Imaging Healthcare). For each subject, the acquisition protocol comprised 64 DWIs at b = 1000 s/mm², along with a pair of b = 0 volumes acquired with reversed phase-encoding directions to facilitate distortion correction. All diffusion data were corrected for susceptibility-induced distortions using FSL's "topup" function[73-75], while eddy-current distortions and subject motion were compensated for using FSL's "eddy" function[76-79].

***WCSUH-Adolescent dataset***

In accordance with local ethics, diffusion MRI data of 0.9×0.9×3.0 mm resolution were acquired from 48 subjects (ages 5–18 years) at West China Second University Hospital (WCSUH). Imaging was performed using a 2D EPI sequence on a Siemens MAGNETOM Skyra 3.0 Tesla scanner. The acquisition protocol comprises 64 b = 1000 s/mm² DWI and two b = 0 volumes. Additionally, a separate b = 0 volume with reversed phase-encoding direction was acquired to facilitate distortion correction. All diffusion data were corrected for susceptibility-induced distortions using FSL's "topup" function, followed by the mitigation of eddy currents and subject motion via FSL's "eddy" function.

***Reference microstructure parameter generation***

To establish the reference data for learning the generative microstructure prior, we estimated parameters for three distinct biophysical models, including the tensor model, the kurtosis model, and the NODDI model. For the tensor model, parameters were derived from ordinary least squares regression implemented in FSL using b = 0 volumes and single-shell data with b-values less than 1500 s/mm² of each scan. Due to the scarcity of high-quality multi-shell diffusion data, the kurtosis model parameters for training were obtained from the HCP-3T dataset, employing ordinary least squares regression in MRtrix3 on all b = 0, b = 1000, and b = 2000 s/mm² data. Similarly, the NODDI model parameters were computed using the NODDI Toolbox on all b = 0, b = 1000, b = 2000, and b = 3000 s/mm² data.

**Out-of-distribution test datasets and analysis**

***CHCP dataset***

Pre-processed diffusion MRI data acquired at an isotropic spatial resolution of 1.5 mm from 21 subjects in the Chinese Human Connectome Project (CHCP) were used for validation across various diffusion encodings and microstructure models. Each acquisition included 14 b = 0 image volumes, 93 b = 1000 s/mm² diffusion-weighted imaging (DWI) volumes, and 92 b = 2000 s/mm² DWI volumes. Reference parameters for the tensor, kurtosis, and NODDI models were computed using all b = 0 volumes and single-shell, two-shell, and two-shell configurations, respectively, and were derived using the same methods applied to the training datasets.

On the CHCP dataset, the gradient table of diffusion MRI was sub-sampled to evaluate PIGMENT's performance in accelerated imaging. For the tensor model, diffusion MRI was sub-sampled to acceleration factors of 14, 8, and 5, corresponding to: (1) 1 b = 0, 6 b = 1000 s/mm² volumes, (2) 2 b = 0, 12 b = 1000 s/mm² volumes, and (3) 3 b = 0, 18 b = 1000 s/mm² volumes. For the NODDI model, diffusion MRI was sub-sampled to acceleration factors of 18, 9, and 6, corresponding to: (1) 1 b = 0, 5 b = 1000, 5 b = 2000 s/mm² volumes, (2) 2 b = 0, 10 b = 1000, 10 b = 2000 s/mm² volumes, and (3) data of 3 b = 0, 15 b = 1000, 15 b = 2000 s/mm² volumes. For the kurtosis model, diffusion MRI was sub-sampled to acceleration factors of 6, 4, and 3, corresponding to: (1) 1 b = 0, 10 b = 1000, 20 b = 2000 s/mm² volumes, (2) 3 b = 0, 15 b = 1000, 30 b = 2000 s/mm² volumes, and (3) data of 4 b = 0, 20 b = 1000, 40 b = 2000 s/mm² volumes.

For the demonstration of ill-posed measurements for DTI, we evaluated extremely sparse gradient sampling schemes. Each scheme included a single b = 0 volume paired with progressively increasing diffusion-weighted directions: (1) one DWIs of left-to-right direction, (2) two DWIs of left-to-right and posterior-to-anterior directions, (3) three DWIs of left-to-right, anterior-to-posterior, and inferior-to-superior directions, (4) four DWIs of tetrahedral diffusion encoding[80], and (5) five DWIs with five uniformly distributed diffusion encoding directions.

For downstream tasks, whole-brain tractography was performed using the clinically-standard fiber assignment by continuous tracking (FACT) algorithm, with FA-modulated V1 as input, implemented in MRtrix3 "tckgen". Anatomically-constrained tractography (ACT) was applied using tissue segmentation derived from a co-registered T1w image, ensuring that streamlines originate and terminate at the gray-white matter interface to improve biological accuracy. Structural connectivity matrices were then constructed using the Desikan-Killiany (DK) atlas, implemented with MRtrix3 "tck2connectome".

***THU-3T dataset***

In accordance with local ethics, pre-processed diffusion MRI data acquired at three spatial resolutions (0.96, 1.5 and 2.0 mm isotropic), together with T1w data (1 mm isotropic) from 20 healthy young adults were used to evaluate PIGMENT's performance across spatial resolutions. All procedures were approved by the institutional review board and written informed consent was obtained from all participants. Data were collected on a Siemens MAGNETOM Prisma 3T scanner with a 32-channel head coil. Submillimeter (0.96 mm isotropic) diffusion data were acquired with the gSlider sequence[81,82], which collected five 4.8-mm-thick slabs with 0.96×0.96 mm$^2$ in-plane resolution using five optimized RF-encoding profiles to reconstruct 0.96 mm-thick slices, achieving 0.96 mm isotropic resolution. Diffusion data at 1.5 and 2.0 mm iso. resolutions were acquired using the vendor-provided simultaneous multi-slice (SMS) pulsed gradient spin-echo (PGSE) single-shot EPI sequence. The diffusion encoding included 32 DWIs at b = 1000 s/mm$^2$ and 64 DWIs at b = 2500 s/mm$^2$ along uniformly distributed directions, with a b = 0 volume inserted every 16 DWIs and an additional reversed-phase b = 0 for susceptibility correction. T1w data were acquired with MPRAGE sequence at 1 mm isotropic resolution.

All images were first corrected for gradient nonlinearity[83]. Diffusion data were then corrected for susceptibility-induced distortions (FSL "topup"), and for eddy-current and motion artifacts (FSL "eddy"). To compensate for the intrinsically low SNR at submillimeter resolutions, the 0.96-mm data were further denoised with BM4D algorithm[84]. Reference tensor at each resolution was fitted using all b = 1000 s/mm$^2$ volumes using ordinary least-squares fitting implemented in FSL "dtifit". White matter masks were generated from each subject's T1w image using FreeSurfer "recon-all" [58] and registered to diffusion space with FreeSurfer "bbregister" [85] for MAE calculation. Full acquisition parameters and processing details are provided in Table S2 and the original dataset publication[86].

***WCSUH-Early-Childhood dataset***

In accordance with local ethics, pre-processed diffusion MRI data of 70 young children aged 0-60 months acquired in WCSUH were used to evaluate the potential of PIGMENT in neuroscience research. Data were collected on a Siemens MAGNETOM Skyra using a product 2D SMS pulsed gradient spin echo (PGSE) single-shot echo planar imaging (EPI) sequence at 0.9 × 0.9 × 3 mm$^3$ resolution, with 2 b = 0 volumes and 64 b = 1000 s/mm$^2$ DWI volumes. Full acquisition parameters are provided in Table S2.

The overall processing pipeline is illustrated in Figure S9. All diffusion MRI data were corrected for eddy current-induced distortions and up-sampled to 0.9 mm isotropic resolution using spline interpolation implemented by FSL "flirt". Six DWIs at b = 1000 s/mm$^2$ were selected to minimize the condition number of the diffusion tensor transformation matrix (i.e., to achieve maximal uniformity) as the accelerated subset, while the reference tensor was estimated using all available b = 1000 s/mm$^2$ volumes.

Two complementary segmentation strategies were employed to define regions of interest. First, for detailed cortical parcellation, the mean DWI was co-registered to an age-appropriate T1w atlas constructed from the BCP dataset[87] using ANTs "SyN" registration with FSL's "fnirt" function serving as a robust alternative in cases of suboptimal registration. The atlas-based volumetric segmentation and cortical parcellation were then transformed to each subject's native space to obtain individual parcels. Second, to isolate the white matter within the frontal lobe, we utilized the FreeSurfer "recon-all-clinical" pipeline with mean DWI as input.

For downstream connectivity analyses, whole-brain tractography was performed using the clinically-standard fiber assignment by continuous tracking (FACT) algorithm, with FA-modulated V1 as input, implemented in MRtrix3 "tckgen". Anatomically-constrained tractography (ACT) was applied based on tissue segmentation, ensuring that streamlines originate and terminate at the gray-white matter interface for greater biological accuracy. The thalamocortical tract was extracted by selecting streamlines with endpoints in both the thalamus and the precentral gyrus. Short-range association fibers (SAFs) were defined as tracts connecting adjacent gyri.

***Oxford-Meso dataset***

In accordance with local ethics, pre-processed diffusion MRI data of 0.7 mm isotropic resolution of one adult acquired at Oxford Centre for Integrative Neuroimaging were used to evaluate PIGMENT for improving ultra-high-resolution microstructural imaging. Data were collected on a Siemens MAGNETOM Prisma scanner using an in-plane segmented 3D multi-slab sequence with inter-volume slab-shifting for slab boundary artifact correction[20,88], comprising four b = 0 volumes and 32 b = 1000 s/mm$^2$ DWI volumes. The total acquisition time was 61 minutes. Comprehensive acquisition and reconstruction details are available at Table S2 and a previous work[20]. Diffusion data were corrected for susceptibility-induced distortions using FSL's "topup", and eddy-current and motion artifacts using FSL's "eddy". The complete dataset of all b = 0 volumes and DWI volumes was used for tensor model fitting and comparative analysis.

The whole-brain segmentation, as well as pial and white matter/gray matter (WM/GM) surface meshes were reconstructed from the same subject's 0.7 mm isotropic T1w MPRAGE image using FreeSurfer "recon-all". Then, cortical columns were generated by connecting corresponding vertices between the pial and WM/GM surfaces, with 21 equidistant points sampled along each column. These anatomical structures were mapped to diffusion space using the transformation derived via FSL "epi_reg".

The ROI-based cortical depth analysis follows a previous pipeline[21]. Specifically, we utilized the "easy_lausanne" version of the Connectome Mapper to implement brain parcellation and obtain 1000 brain ROIs[89]. The cortical depth profiles of FA, MD, and radial index (RI) of single cortical columns were computed by averaging in these ROIs.

***UCSF-0.55T dataset***

In accordance with local ethics, low-field diffusion MRI data were acquired from 10 subjects at a spatial resolution of 1.9×1.9×3.0 mm using a 2D echo planar imaging (EPI) sequence on the 0.55-Tesla MAGNETOM Free.Max scanner (Siemens). To derive high-quality tensor model parameters, we collected four repetitions of 1 b = 0 volume and 6 b = 1000 s/mm$^2$ DWI volumes, followed by four repetitions of 1 b = 0 volume and 12 b = 1000 s/mm$^2$ DWI volumes. This approach enhanced the DWI's SNR of each direction. The susceptibility-induced distortions were corrected using FSL's "topup" function. For diffusion analysis, each volume was up-sampled to 1.5 mm isotropic resolution using spline interpolation via FSL's "flirt" function. The reference tensor model parameters were generated using all 8 b = 0 volumes and 72 b = 1000 s/mm$^2$ DWI volumes. Additionally, a T1w image volume was acquired for each subject using the MPRAGE sequence at a spatial resolution of 1.0×1.0×1.0 mm. Whole-brain segmentation was generated from the T1w volume using the "recon-all" function of FreeSurfer software for each subject.

***Ghent-Tumor dataset***

Diffusion MRI and T1w data, along with tumor masks, from 25 patients (14 WHO Grade I-II meningiomas and 11 WHO Grade II-III gliomas) in a publicly available dataset [90] were used to evaluate the generalization capability and utility of PIGMENT under pathological conditions. Data were collected on a Siemens MAGNETOM Trio scanner with a 32-channel head coil in Ghent University Hospital. Diffusion data were acquired at 2.5 mm isotropic resolution using a multi-shell protocol (6 b = 0 volumes, 16 b = 700 s/mm$^2$, 30 b = 1200 s/mm$^2$, and 50 b = 2000 s/mm$^2$ DWIs), with an additional reversed-phase b = 0 volume acquired to facilitate susceptibility correction. T1w data were acquired using an MPRAGE sequence at 1 mm isotropic resolution. Full acquisition parameters are provided in Table S2 and the original dataset publication[90]. Diffusion data were corrected for susceptibility-induced distortions (FSL "topup"), eddy-current and motion artifacts (FSL "eddy"), and subsequently up-sampled to 1.5 mm isotropic resolution for analysis with bicubic interpolation.

To construct the accelerated subsets, six DWIs that minimizes the condition number of the diffusion tensor transformation matrix (i.e., to achieve maximal uniformity) and three DWIs along three orthogonal directions at b = 1200 s/mm$^2$ were selected respectively. Meanwhile the reference tensor was estimated using all available b = 1200 s/mm$^2$ volumes. Tissue segmentations were generated from T1w images using FreeSurfer "recon-all" and registered to the 1.5 mm isotropic diffusion space using FreeSurfer "bbregister". Tumor masks were registered to the same space using ANTs rigid transform. These aligned masks were then utilized to explicitly exclude pathological tissues from the normal gray matter (GM), white matter (WM), and cerebrospinal fluid (CSF) segmentations.

For the downstream white matter pathway reconstruction, whole-brain tractography was performed using the clinically standard fiber assignment by continuous tracking (FACT) algorithm, based on FA-modulated V1, was employed using MRtrix3 "tckgen" with anatomically constrained tractography (ACT) to ensure that streamlines originated and terminated at the gray-white matter interface, thereby enhancing biological accuracy. The superior longitudinal fasciculus (SLF) was dissected by selecting tracts penetrating both SLF way-point ROIs, which were defined by registering the JHU atlas to the individual diffusion space using ANTs "SyN" transform.

***WCH-Tumor dataset***

In accordance with local ethics, pre-processed diffusion MRI data from six patients with glioma from West China Hospital of Sichuan University (WCH) were used to evaluate the capability for mapping brain microstructure from ill-posed acquisitions (i.e., 3 DWIs). Data were collected on a Siemens MAGNETOM Skyra scanner. Diffusion data were acquired at $0.9 \times 0.9 \times 3$ mm$^3$ resolution using a single-shell protocol (2 b = 0 volumes, 60 b = 1000 s/mm$^2$ DWIs). Full acquisition parameters are provided in Table S2.

Diffusion data were corrected for eddy-current and motion artifacts (FSL "eddy"), and subsequently up-sampled to 1.5 mm isotropic resolution using bicubic interpolation. To simulate a rapid clinical protocol, a subset of one b = 0 and DWIs along three orthogonal axes (i.e., [1, 0, 0], [0, 1, 0], and [0, 0, 1]) at b = 1000 s/mm$^2$ were selected.

Whole-brain tractography was generated using the clinically-standard fiber assignment by continuous tracking (FACT) algorithm based on FA-modulated V1, implemented in MRtrix3 "tckgen", in combination with ACT to ensure that streamlines originated and terminated at the gray-white matter interface, thereby enhancing biological accuracy. Tissue segmentation derived from the b = 0 image using "mri_synthseg" was corrected prior to ACT by excluding tumor regions from the gray matter, white matter, and CSF masks.

**Diffusion modeling**

Tensor model characterizes water diffusion using a second-order symmetric tensor $[D_{11}\ D_{22}\ D_{33}\ D_{12}\ D_{13}\ D_{23}\ ]^T$, where $D_{jk}$ ($j$, $k$ = 1, 2, 3) represents the six unique elements of the diffusion tensor. The modeling process needs to estimate the symmetric tensor and non-weighted signal value as $\boldsymbol{P} = [D_{11}\ D_{22}\ D_{33}\ D_{12}\ D_{13}\ D_{23}\ S_0]^T$. The forward process to synthesize signal from estimated parameters along $i^{th}$ direction with the b-value $b_i$ and the gradient encoding direction $\boldsymbol{v_i}$ is:

$$S_i = tensor(\boldsymbol{P}, \boldsymbol{v_i}, b_i) = S_0 e^{-b_i \sum_{j=1}^{3} \sum_{k=1}^{3} v_{ij} v_{ik} D_{jk}}, \quad (14)$$

Kurtosis model is an advanced extension of the tensor model that captures non-Gaussian water diffusion properties. It has additional fourth-order three-dimensional tensor with 15 independent components $\boldsymbol{K}$ that characterizes the complexity of tissue microstructure compared to tensor model's six components $\boldsymbol{D}$. The modeling process needs to estimate totally 22 parameters $\boldsymbol{P} = [\boldsymbol{D^T}\ \boldsymbol{K^T}\ S_0]^{\boldsymbol{T}}$ from acquired data. The forward process to synthesize signal from estimated parameters along $i^{th}$ direction with the b-value $b_i$ and the gradient encoding direction $\boldsymbol{v_i}$ is:

$$S_i = kurtosis(\boldsymbol{P}, \boldsymbol{v_i}, b_i) = S_0 e^{-b_i \sum_{j=1}^{3} \sum_{k=1}^{3} v_{ij} v_{ik} D_{jk} + \frac{1}{6} {b_i}^2 \left(\frac{1}{3} \sum_{j=1}^{3} D_{jj}\right)^2 K_i^{app}}, \quad (15)$$

$$K_i^{app} = \sum_{j=1}^{3}\sum_{k=1}^{3}\sum_{l=1}^{3}\sum_{m=1}^{3} v_{ij}v_{ik}v_{il}v_{im}K_{jklm}\,, \tag{16}$$

where $D_{jk} = D_{kj}$ and $K_{jklm} = K_{sort(j,k,l,m)}$ due to the symmetric properties of $\boldsymbol{D}$ and $\boldsymbol{K}$ matrixes.

NODDI model is an advanced biophysical model for diffusion MRI that characterizes brain tissue microstructure using six parameters $\boldsymbol{P} = [S_0, f_{iso}, f_{ic}, \boldsymbol{\mu}^{\boldsymbol{T}}, \kappa]^T$ by disentangling water diffusion into intra-cellular compartment $A_{ic}(\boldsymbol{\mu}, \kappa, \boldsymbol{v_i}, b_i)$ , extra-cellular compartment $A_{ec}(\boldsymbol{\mu}, f_{ic}, \boldsymbol{v_i}, b_i)$ and cerebrospinal fluid (CSF) compartment $A_{iso}$. The modeling process needs to estimate the parameters $\boldsymbol{P}$ from acquired signals including non-diffusion-weighted signal value ($S_0$), isotropic volume fraction ($f_{iso}$, ISOVF), intra-cellular volume fraction ($f_{ic}$, ICVF), mean orientation ($\boldsymbol{\mu}$, a three-dimensional vector on the unit sphere), and orientation dispersion index (ODI) derived from $\kappa$. The forward process to synthesize signal from estimated parameters is:

$$S_i = NODDI(\boldsymbol{P}, \boldsymbol{v_i}, b_i, S_0) = S_0\big((1 - f_{iso})(f_{ic}A_{ic} + (1 - f_{ic})A_{ec}) + f_{iso}A_{iso}\big), \tag{17}$$

where $b_i$ and $\boldsymbol{v_i}$ represents the b-value and the gradient encoding direction of $i^{th}$ data volume.

**Quantification metrics**

The microstructure parameter estimation quality was quantified using the mean absolute error (MAE) for tensor, NODDI, and kurtosis model metrics. However, specifically for the kurtosis measures derived from kurtosis model (i.e., axial kurtosis and radial kurtosis), the non-linear fitting process is prone to numerical instability, often yielding outliers that distort mean-based statistics. To mitigate this, we adopted the median absolute error as a robust alternative to evaluate the accuracy of these specific kurtosis parameters. The relative performance improvement was calculated in relation to the conventional method (con. method) utilized in the analysis pipeline, which is ordinary least squared regression implemented in FSL for tensor model, point-wise grid search and gradient descent implemented in NODDI-Toolbox[91] for NODDI model, and ordinary least squared regression implemented in MRtrix3 for kurtosis model:

$$\text{Improvement}(method) = \frac{MAE(con.\ method) - MAE(method)}{MAE(con.\ method)} \times 100\%. \tag{18}$$

Tensor model metrics included fractional anisotropy (FA), mean diffusivity (MD), axial diffusivity (AD) and primary eigenvector (V1). NODDI model metrics included isotropic volume fraction (ISOVF), intra-cellular volume fraction (ICVF) and orientation dispersion index (ODI). Kurtosis model metrics included fractional anisotropy (FA), axial kurtosis (AK), and radial kurtosis (RK).

**Network implementation**

All neural networks were implemented using the PyTorch platform[92]. The autoencoder that maps diffusion model parameters into latent space in Fig. 1a was a 3D vector quantized model with a $k$-channel ($k$ = 7, 7, and 22 for tensor model, NODDI model, and kurtosis model, respectively) latent embedding dimension and a codebook size of 16,000, featuring three down-sampling layers and two residual blocks at each scale. The denoising network Unet $\epsilon_\theta$ in Fig. 1a was implemented as an attention U-Net operating across four resolution scales. At each scale, a self-attention module was positioned between two residual blocks. During training, each input volume was randomly split into 64×64×64×channel blocks with flipping along the anatomical left-right axis for data augmentation. Network parameters were optimized using Adam optimizer with learning rate of $5\times10^{-5}$ for the autoencoder and $2.5\times10^{-5}$ for the latent space denoiser, with a batch size of 12 on two NVIDIA A800 GPUs. During inference, the acquired DWI volume was divided into blocks of size 64×64×64×channels with minimal overlap for processing, with a total of 500 DDIM steps employed. The dual domain optimization was performed at the $480^{th}$ and $500^{th}$ steps using a learning rate of $1\times10^{-5}$.

**Baseline methods**

For the tensor model, the baseline methods included OLS[56,93,94], DTI-RIM[40], SH-DTI[29], and DIMOND[32]. The OLS implemented in FSL's "dtifit" function is the conventional approach for fitting tensor model parameters. DTI-RIM utilizes tensor model parameter predictions and their gradient as state variables, and employed a recurrent inference machine (RIM) to predict the subsequent more accurate state. DTI-RIM has demonstrated robust generalization across diverse q-space sampling pattern. DIMOND's implementation here utilized tensor model for signal synthesis.

For the NODDI model, the baseline methods included NODDI-Toolbox[91], SH-NODDI[31] and DIMOND[32]. NODDI-Toolbox is conventional software to solve NODDI model parameters which employed a point-wise grid search and gradient descent. SH-NODDI employs a 15-layer 3D UNet with four down-sampling layers to map the combined spherical harmonic representations fitted from each b-value to NODDI model metrics, including ISOVF, ICVF, and ODI and was trained on HCP dataset. Due to the limited diffusion signals contained in each b-value (maximum of 15), we utilized second-order spherical harmonic functions. DIMOND is a self-supervised method that employed a seven-layer plain 3D convolutional neural network (CNN) to map each subject's input diffusion data to diffusion model parameters. The network parameters are optimized by minimizing the discrepancy between the acquired signal and the synthetic signal generated through the forward diffusion model using neural network predictions.

For the diffusion kurtosis model, the baseline estimators included ordinary least squares (OLS), iterative weighted least squares (IWLS)[57], and constrained weighted least squares (CWLS) regression[95]. Standard OLS, implemented via MRtrix3, served as the conventional baseline for parameter estimation. To mitigate noise effects, we evaluated IWLS (also implemented in MRtrix3), which iteratively applies weights to the L2 loss function between the acquired and model-predicted signals. Furthermore, we included the CWLS method, implemented in the DESIGNER software package. This approach builds upon the WLS framework by enforcing non-negativity constraints on the apparent excess kurtosis, thereby suppressing physically implausible outliers in parameter estimation.

## Acknowledgments

Preliminary results of this research were submitted on November 11th, 2024, to the annual scientific meeting of the International Society of Magnetic Resonance in Medicine (ISMRM) and were accepted for the oral presentation in May 2025, Honolulu.

### Funding:

This work was supported by the following fundings:
National Natural Science Foundation of China 82302166 (Q.T.)
Tsinghua University Dushi Project and Startup Fund (Q.T.)
Beijing Natural Science Foundation QY24283 (J. Z.)

### Competing interests:

The authors declare the following competing interests: Q.T., Z. L., J. Z., H. Y., M. L., H. L. are named as inventors on a granted patent related to the research described in this manuscript (Certificate No. CN120471881B, entitled "Method, apparatus, and electronic device for microstructure model optimization based on diffusion prior"). All other authors declare no competing financial or non-financial interests.

### Data, code, and materials availability:

Code and data will be made publicly available upon acceptance of the manuscript.

The diffusion data were provided by the Human Connectome Project, WU-Minn-Ox Consortium (Principal Investigators: David Van Essen and Kamil Ugurbil; U54-MH091657) funded by the 16 NIH Institutes and Centers that support the NIH Blueprint for Neuroscience Research; and by the McDonnell Center for Systems Neuroscience at Washington University. Data were provided [in part] by the Chinese Human Connectome Project (CHCP, PI: Jia-Hong Gao) funded by the Beijing Municipal Science & Technology Commission, Chinese Institute for Brain Research (Beijing), National Natural Science Foundation of China, and the Ministry of Science and Technology of China.

Data used in the preparation of this article were obtained from the Alzheimer's Disease Neuroimaging Initiative (ADNI) database (adni.loni.usc.edu). The ADNI was launched in 2003 as a public-private partnership, led by Principal Investigator Michael W. Weiner, MD. The primary goal of ADNI has been to test whether serial magnetic resonance imaging (MRI), positron emission tomography (PET), other biological markers, and clinical and neuropsychological assessment can be combined to measure the progression of mild cognitive impairment (MCI) and early Alzheimer's disease (AD).

Data collection and sharing for the Alzheimer's Disease Neuroimaging Initiative (ADNI) is funded by the National Institute on Aging (National Institutes of Health Grant U19 AG024904). The grantee organization is the Northern California Institute for Research and Education. In the past, ADNI has also received funding from the National Institute of Biomedical Imaging and Bioengineering, the Canadian Institutes of Health Research, and private sector contributions through the Foundation for the National Institutes of Health (FNIH) including generous contributions from the following: AbbVie, Alzheimer's Association; Alzheimer's Drug Discovery Foundation; Araclon Biotech; BioClinica, Inc.; Biogen; Bristol-Myers Squibb Company; CereSpir, Inc.; Cogstate; Eisai Inc.; Elan Pharmaceuticals, Inc.; Eli Lilly and Company; EuroImmun; F. Hoffmann-La Roche Ltd and its affiliated company Genentech, Inc.; Fujirebio; GE Healthcare; IXICO Ltd.; Janssen Alzheimer Immunotherapy Research & Development, LLC.; Johnson & Johnson

Pharmaceutical Research &Development LLC.; Lumosity; Lundbeck; Merck & Co., Inc.; Meso Scale Diagnostics, LLC.; NeuroRx Research; Neurotrack Technologies; Novartis Pharmaceuticals Corporation; Pfizer Inc.; Piramal Imaging; Servier; Takeda Pharmaceutical Company; and Transition Therapeutics.

Data used in the preparation of this article was obtained on [2025-07-11] from the Parkinson's Progression Markers Initiative (PPMI) database (www.ppmi-info.org/access-data-specimens/download-data), RRID:SCR_006431. For up-to-date information on the study, visit www.ppmi-info.org. PPMI – a public-private partnership – is funded by the Michael J. Fox Foundation for Parkinson's Research and funding partners, including 4D Pharma, Abbvie, AcureX, Allergan, Amathus Therapeutics, Aligning Science Across Parkinson's, AskBio, Avid Radiopharmaceuticals, BIAL, BioArctic, Biogen, Biohaven, BioLegend, BlueRock Therapeutics, Bristol-Myers Squibb, Calico Labs, Capsida Biotherapeutics, Celgene, Cerevel Therapeutics, Coave Therapeutics, DaCapo Brainscience, Denali, Edmond J. Safra Foundation, Eli Lilly, Gain Therapeutics, GE HealthCare, Genentech, GSK, Golub Capital, Handl Therapeutics, Insitro, Jazz Pharmaceuticals, Johnson & Johnson Innovative Medicine, Lundbeck, Merck, Meso Scale Discovery, Mission Therapeutics, Neurocrine Biosciences, Neuron23, Neuropore, Pfizer, Piramal, Prevail Therapeutics, Roche, Sanofi, Servier, Sun Pharma Advanced Research Company, Takeda, Teva, UCB, Vanqua Bio, Verily, Voyager Therapeutics, the Weston Family Foundation and Yumanity Therapeutics.

We gratefully acknowledge the Center for Biomedical Imaging Research at Tsinghua University for their pivotal support in data acquisition.

## Supplementary Materials

Supplementary Figs. 1-9
Supplementary Tables 1-2
Source Data